\documentclass[11pt]{article}
\usepackage[a4paper,margin=1.85cm]{geometry}
\usepackage[T1]{fontenc}
\usepackage[utf8]{inputenc}
\usepackage{microtype,setspace,graphicx,booktabs,longtable,tabularx,array,float,amsmath,amssymb,xcolor,caption,fancyhdr,hyperref,siunitx}
\usepackage[backend=biber,style=authoryear,maxcitenames=2,maxbibnames=99]{biblatex}
\hypersetup{colorlinks=true,linkcolor=black,citecolor=black,urlcolor=blue}
\definecolor{navy}{HTML}{23395B}
\definecolor{teal}{HTML}{406E8E}
\definecolor{green}{HTML}{4C956C}
\definecolor{gold}{HTML}{D9B44A}
\title{\textbf{A Path-Dependent Agent-Based Microsimulation of Crime and Violence Reduction Policy Portfolios in Bolivia}\\[0.35em]\large Survey Calibration, Adaptive Emulator Ensembles, Global Sensitivity, Tempered MCMC, and Multi-Objective Policy Analysis}
\author{Ricardo Alonzo Fernández Salguero}
\date{October 2026}
\begin{document}
\maketitle

\begin{abstract}
This study constructs a reproducible policy laboratory for crime and violence reduction in Bolivia using a path-dependent agent-based microsimulation calibrated to the 2025 Household Survey and the fourth-quarter 2025 Continuous Employment Survey. The objective is not to forecast crime mechanically or to identify a single universally superior intervention. It is to represent multiple mechanisms through which victimization, offending, severe violence, reporting, institutional legitimacy, spatial displacement, criminal adaptation, and policy capacity can evolve jointly, and then to compare portfolios under explicit structural and calibration uncertainty. The model combines a survey-weighted synthetic population created from official expansion factors and cross-survey donor matching, heterogeneous routine exposure, prior victimization and offending, socioeconomic shocks, place risk, illicit-network pressure, community organization, police and justice capacity, municipal instruments, social programs, and criminal adaptation. Policy modules include hot-spots policing, problem-oriented policing, focused deterrence, procedural justice, community social control, Guardabosques-type conditional collective incentives, youth and adult behavioral interventions, employment and school-retention measures, street lighting and place remediation, alcohol regulation, violence interruption, witness protection, homicide investigation, financial-network disruption, rehabilitation, and oversight. Empirical targets are kept separate from external intervention priors and structural assumptions. The baseline ABM ensemble is emulated by an adaptive stack combining Gaussian processes with a Bayesian-bootstrap tree ensemble; global sensitivity is examined with Sobol and active-subspace methods; baseline structural parameters are calibrated using a tempered ensemble MCMC with adaptive covariance preconditioning; and policy scenarios are propagated through posterior, intervention-prior, and structural uncertainty. In the included development run, the 2025 Household Survey implies a weighted 12-month victimization rate of 13.65\%, formal reporting of 16.91\% among victims, normalized police trust of 0.470, and normalized perceived safety of 0.488. The fully integrated package reduces simulated victimization by 17.9\% relative to a paired baseline, with a 5th--95th percentile range of approximately 15.1--20.9\%; focused place-based packages achieve similar but lower-cost reductions while enforcement-heavy portfolios generate larger displacement. Community-control and Guardabosques-type modules produce smaller national-average effects because the model deliberately treats their strongest evidence as local and conditional rather than universal. These figures are model-conditional scenario outputs, not causal estimates for Bolivia. The principal contribution is an auditable architecture for testing combinations, trade-offs, and failure modes before policy scale-up.
\end{abstract}

\noindent\textbf{How to cite:} Fernández Salguero, R. A. (2026). \textit{A Path-Dependent Agent-Based Microsimulation of Crime and Violence Reduction Policy Portfolios in Bolivia: Survey Calibration, Adaptive Emulator Ensembles, Global Sensitivity, Tempered MCMC, and Multi-Objective Policy Analysis}. Zenodo. \href{https://doi.org/10.5281/zenodo.23093826}{https://doi.org/10.5281/zenodo.23093826}.

\section{Research problem, evidence base, and empirical calibration}

Crime and violence policy is unusually vulnerable to category errors. A decline in arrests is not necessarily a decline in crime; an increase in recorded victimization can reflect either worsening security or improved reporting; a locally successful enforcement operation can displace activity to nearby places; a community program can reduce conflict while leaving organized markets intact; and an intervention that appears effective in a small high-risk population may have a negligible national-average effect. These problems become more severe when several interventions operate simultaneously. The relevant policy question is therefore not simply whether a particular program ``works'', but how multiple mechanisms interact under capacity constraints, heterogeneous exposure, behavioral adaptation, uncertainty in causal transportability, and competing objectives such as victimization reduction, severe-violence reduction, legitimacy, rights, fiscal cost, and displacement.

The evidence review that motivates this model does not support an undifferentiated enforcement strategy. Meta-analytic evidence for hot-spots policing, problem-oriented policing, and focused deterrence is comparatively strong, but the effects are concentrated in particular places, problems, or high-harm groups rather than being uniform population treatments \parencite{turchan2024,hinkle2020,braga2026}. Community policing is more heterogeneous, including coordinated Global South experiments that did not produce average improvements in either crime or trust \parencite{blair2021}. Procedural-justice interventions are more consistently linked to legitimacy and cooperation than to immediate violent-crime reductions. Environmental interventions such as improved street lighting reduce selected crime outcomes but do not behave as universal violence treatments \parencite{welsh2022,macdonald2026lighting}. Cognitive-behavioral, restorative, youth, reentry, and related social interventions operate through different horizons and populations \parencite{sherman2015,lipsey2007,park2008}. The preceding critical review archived at Zenodo DOI 10.5281/zenodo.23088877 is used in this package as an external policy-prior evidence matrix; the present simulation does not treat that review as a source of fixed effect sizes.

Bolivia contributes two kinds of empirical information. The first is socioeconomic structure. The fourth-quarter 2025 Continuous Employment Survey contains 52,650 observations and 121 variables in the supplied public-use archive. Department, area, household type, education, labor-force status, occupation, industry, working hours, labor income, and expansion factors are used to define labor-market heterogeneity and the demographic-economic environment of synthetic agents \parencite{INE_ECE2025}. The second is household and security information from the 2025 Household Survey. Seven supplied SPSS modules cover persons, housing, equipment, food expenditures, non-food expenditures, food insecurity, and discrimination/security. Crucially, the discrimination module also contains questions on perceived safety, victimization during the previous twelve months, formal reporting, and confidence in the Bolivian Police. Those variables make it possible to anchor baseline victimization, reporting, and institutional trust directly to survey evidence instead of importing them from another country \parencite{INE_EH2025}.

The package contains a pure-Python reader for the compressed SPSS system-file format because the execution environment did not provide \texttt{pyreadstat}. This is not merely a convenience layer: it allows the public-use EH files to be read without altering the original archives and preserves value labels needed to interpret the security module. The parser was validated against the known EH person-file header of 39,485 cases and a nominal case size of 531 storage slots. The discrimination module contains 12,358 observations. Its value labels identify \texttt{S09B\_01} as a four-category perceived-safety item; \texttt{S09B\_02A} and \texttt{S09B\_02B} as first and second victimization categories; \texttt{S09B\_03} as formal reporting; and \texttt{S09B\_04} as confidence in the Bolivian Police. Victimization categories include street theft or robbery, burglary of a dwelling or business, vehicle theft, fraud or abuse of trust, assault or injury, domestic violence, and other crime.

Survey quantities are transformed cautiously. The weighted victimization target equals the probability that a respondent reports at least one category from 1 through 7 in either victimization slot. The reporting target is the weighted share with a formal report among respondents classified as victims. Police trust is normalized to the unit interval after excluding the no-answer category, with higher values representing greater trust. Perceived safety is similarly normalized from the four ordered categories. Poverty and extreme poverty use the survey's own income-poverty indicators. The food-insecurity quantity used in the model is deliberately broad: it is an ``any affirmative'' indicator across the eight food-security questions rather than an official graded severity classification. That quantity is therefore a vulnerability covariate rather than an official prevalence estimate of severe food insecurity. ECE labor-force weights are used for the urban share, open unemployment among the economically active population, and the median labor income. These definitions are reproduced in the data dictionary and source code.

\begin{table}[H]\centering\small\caption{Survey-calibrated targets used by the microsimulation}\begin{tabularx}{\textwidth}{Xrr}\toprule Indicator & Value & Source\\\midrule
Urban population share & 0.694 & ECE 4T-2025\\
Open unemployment among economically active population & 0.017 & ECE 4T-2025\\
Median monthly labor income (Bs) & 2,500.0 & ECE 4T-2025\\
Income poverty rate & 0.447 & EH-2025\\
Extreme income poverty rate & 0.167 & EH-2025\\
Median household per-capita monthly income (Bs) & 1,347.1 & EH-2025\\
12-month victimization rate (listed EH categories) & 0.136 & EH-2025\\
Formal reporting rate conditional on victimization & 0.169 & EH-2025\\
Police trust score normalized to [0,1] & 0.470 & EH-2025\\
Perceived safety score normalized to [0,1] & 0.488 & EH-2025\\
Broad any-affirmative food-insecurity indicator & 0.543 & EH-2025\\
\bottomrule\end{tabularx}\end{table}

The empirical targets are informative about baseline exposure and vulnerability, but they do not by themselves identify the causal effect of policing, community monitoring, lighting, behavioral treatment, or organized-crime investigation. The model therefore maintains a strict provenance distinction. Survey targets are tagged as \emph{survey-calibrated}. Intervention-effect distributions are tagged as \emph{external evidence}. Criminal adaptation, displacement, institutional capture, and the severity multiplier are tagged as \emph{structural assumptions}. The baseline hazard, reporting shift, and trust drift are tagged as \emph{calibrated} once updated by the likelihood. Scenario outcomes are tagged as \emph{simulated}. This separation is essential because otherwise a calibrated model could create the false impression that every structural relationship has been estimated from Bolivian microdata.

The weighted EH targets indicate a 12-month victimization rate of 0.1365, a formal reporting rate of 0.1691 among victims, normalized police trust of 0.4703, and normalized perceived safety of 0.4883. The weighted income-poverty rate is 0.4468 and the extreme-poverty rate is 0.1669. The ECE urban share is 0.6940, and median monthly labor income is Bs 2,500. The broad food-insecurity vulnerability indicator is 0.5434. The open unemployment target obtained from the supplied ECE coding is 0.0170. That last quantity is used only as a labor-market state target under the survey's coding and should not be read as a claim about alternative unemployment definitions.

\begin{figure}[H]\centering
\includegraphics[width=.74\textwidth]{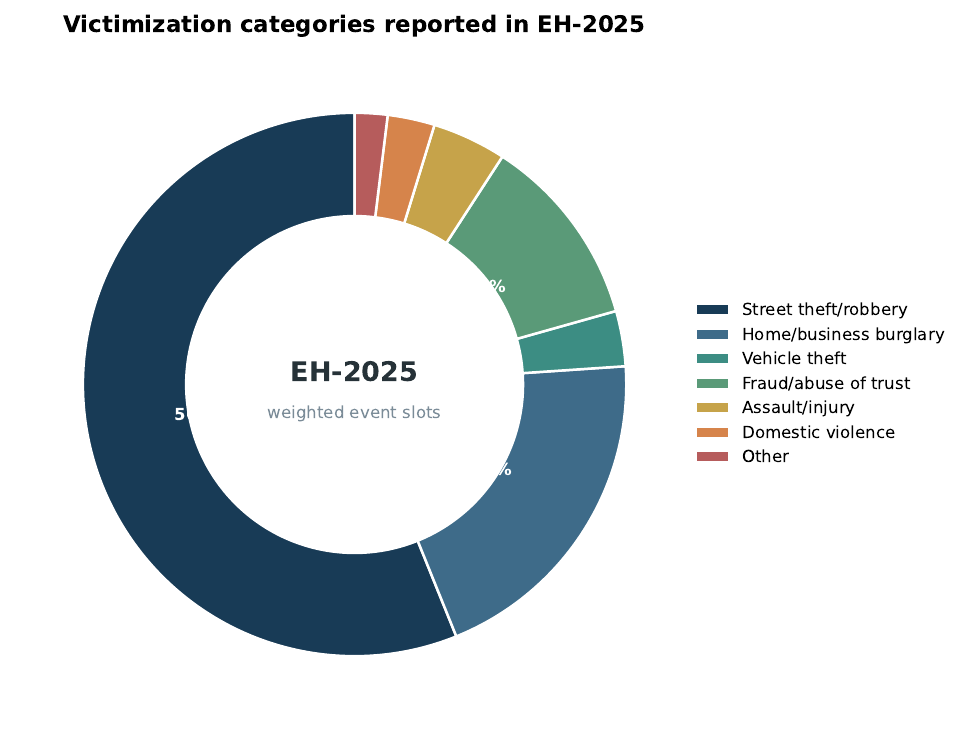}
\caption{Weighted composition of EH-2025 victimization event slots among listed categories 1--7. A respondent can contribute two event categories, so shares describe event-slot composition rather than mutually exclusive population prevalence.}\label{fig:victimdonut}
\end{figure}

Figure~\ref{fig:victimdonut} uses the weighted event slots of the EH victimization items to describe the composition of listed victimization among categories 1--7. Street theft or robbery represents the largest weighted share of event slots, followed by home or business burglary and fraud or abuse of trust. This descriptive composition is not a prevalence decomposition because one respondent can report two categories. It is nevertheless useful for deciding which crime channels deserve explicit representation in the ABM.

\begin{figure}[H]\centering
\includegraphics[width=.84\textwidth]{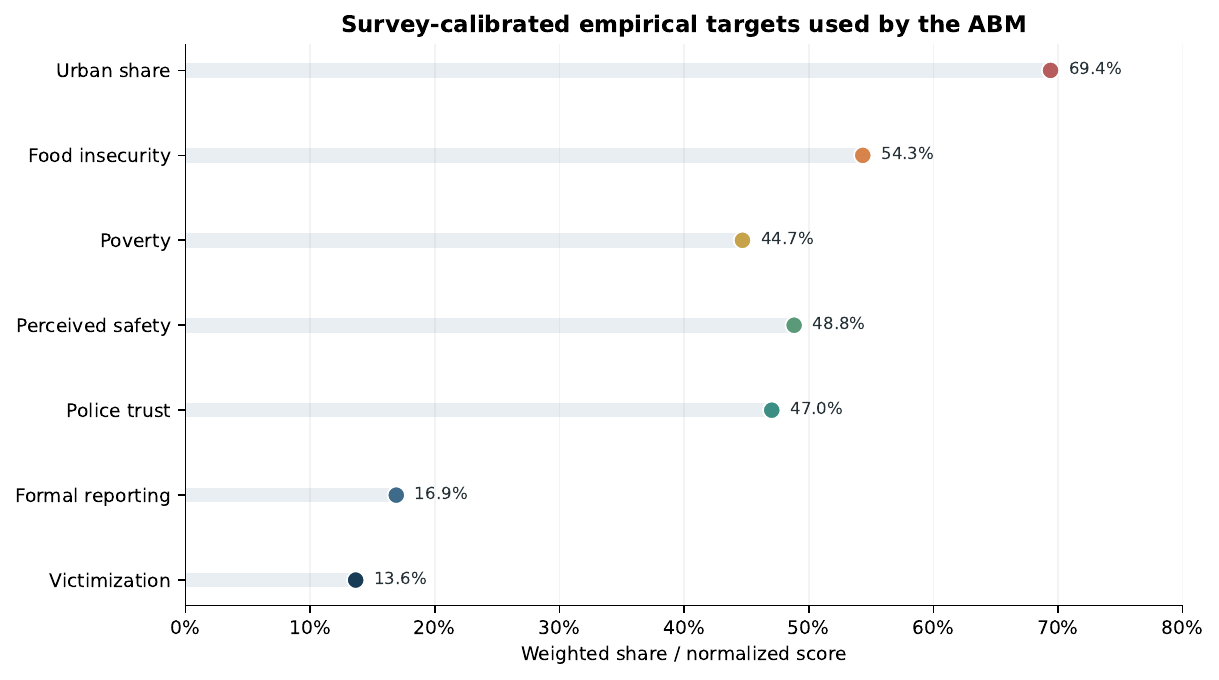}
\caption{Empirical calibration targets. Shares and normalized trust/safety scores are displayed on a common 0--1 axis; income quantities are reported separately in Table 1.}\label{fig:targets}
\end{figure}

The research questions are consequently comparative and mechanism-specific. First, which combinations reduce simulated victimization and severe violence once common-random-number pairing, posterior calibration uncertainty, intervention-prior uncertainty, displacement, capture, and adaptation are propagated? Second, how do place-based enforcement, focused deterrence, community social control, Guardabosques-type collective incentives, social prevention, environmental design, justice legitimacy, investigation, and financial-network disruption differ in the outcomes they affect? Third, what policy channels dominate the global sensitivity of the simulated victimization response? Fourth, do integrated packages produce materially different results across poverty, urban/rural, and age groups in a distributional stress test? Fifth, which scenario portfolios remain non-dominated once resource cost and displacement are included instead of judging policies solely by their point estimate for victimization?

\section{Path-dependent agent architecture and crime mechanisms}

The simulation is organized as a hierarchical microsystem rather than as a single representative offender or representative neighborhood. The synthetic population is generated through survey-weighted household/person resampling and cross-survey donor fusion, not independent marginal Bernoulli draws. Persons remain nested in synthetic households, are assigned to departments and urban or rural areas, and carry joint demographic, education, labor, poverty, food-security, victimization, trust, and perceived-safety traits inherited from weighted survey donors before being exposed to heterogeneous place risks. The full conceptual architecture contains person, household, business, school, community organization, police unit, prosecutor/court, municipal government, social program, illicit market, criminal organization, and place states. The implementation vectorizes person-level state for computational efficiency while preserving typed interfaces for institutional and territorial agents. This design avoids the computational overhead of treating every object as a Python class while retaining explicit state boundaries and policy channels.

\begin{figure}[H]\centering
\includegraphics[width=.98\textwidth]{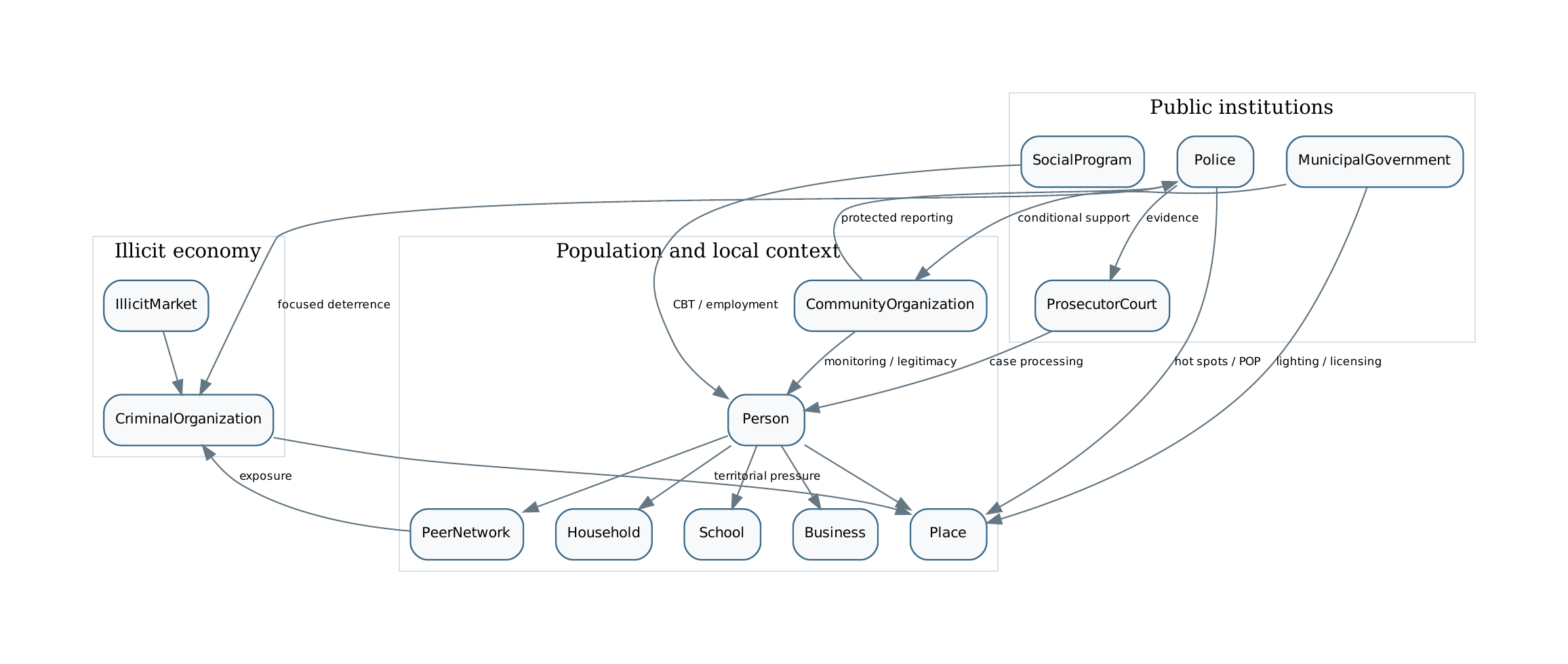}
\caption{Conceptual interaction architecture. The implementation vectorizes person and place states but preserves separate institutional mechanisms.}\label{fig:agents}
\end{figure}

The person state contains age, household membership, income, employment status, poverty, food insecurity, baseline victimization, institutional trust, and a latent risk index. The risk index does not define criminality by demographic identity. It is an evolving state affected by economic stress, prior victimization, prior offending, social interventions, community monitoring, and a stochastic innovation. Households pool vulnerability and guardianship conditions. Places contain baseline risk, lighting, alcohol exposure, and guardianship. Community organizations contain legitimacy, monitoring capacity, and capture risk. Police and justice institutions have capacity, procedural quality, integrity, evidence quality, and workload concepts. Criminal organizations and illicit markets contain coercive capacity, profitability, displacement potential, and adaptation. Social programs have capacity and implementation fidelity. These entities interact through the pathways shown in Figure~\ref{fig:agents}.

Crime is generated through competing event channels rather than a single crime probability. For individual $i$ in place $j$, event type $k$, and month $t$, the conceptual latent hazard is
\begin{equation}
\lambda_{ijkt}=\lambda_{0k}\exp\left(X_{it}\beta_k+H_{ht}\eta_k+P_{jt}\gamma_k+N_{it}\nu_k+M_{jt}\mu_k+\Pi_{ijkt}\theta_k+\xi_{ijkt}\right),
\end{equation}
where $X$ is individual state, $H$ household vulnerability and guardianship, $P$ place risk, $N$ peer or criminal-network exposure, $M$ illicit-market pressure, and $\Pi$ policy exposure. The development implementation uses a bounded numerical version of this formulation so probabilities remain finite. Six event classes are explicit: opportunistic property crime, robbery, interpersonal violence, domestic or relationship violence, extortion/coercion, and organized illicit-market violence. The event channels can overlap in a month, which matters when interpreting aggregate counts. The baseline victimization target is therefore matched through calibration rather than by forcing one channel to equal the survey prevalence.

Path dependence enters through several routes. Prior victimization increases future exposure by affecting routines, neighborhood attachment, retaliation risk, and latent vulnerability. Prior offending increases subsequent offending propensity but decays over time and can be reduced by behavioral or reintegration programs. Place risk persists but can be altered by lighting, guardianship, problem-oriented responses, or displacement. Trust changes after reporting, unreported violence, procedural-justice exposure, and the calibrated trust drift. Criminal adaptation moves risk when enforcement concentrates on specific places or networks. Community capture weakens community-control and Guardabosques-type mechanisms. Reentry and rehabilitation alter the post-sanction trajectory rather than assuming that incapacitation permanently removes risk.

The model contains a bounded expected-offending component,
\begin{equation}
U^{off}_{it}=G_{it}-p^{detect}_{it}C^{sanction}_{it}-C^{moral}_{it}-C^{opportunity}_{it}+\rho N^{peer}_{it}+\omega S^{shock}_{it}+\varepsilon_{it},
\end{equation}
but does not equate behavior with deterministic rational choice. The stochastic hazard incorporates incomplete information, social influence, bounded rationality, impulsivity, and unobserved shocks. A person can move into or out of higher-risk states. No sex, ethnicity, poverty category, or other demographic attribute mechanically makes a person an offender. Socioeconomic variables affect exposure and vulnerability probabilistically and are never interpreted as moral or biological determinants of criminality.

\begin{figure}[H]\centering
\includegraphics[width=.98\textwidth]{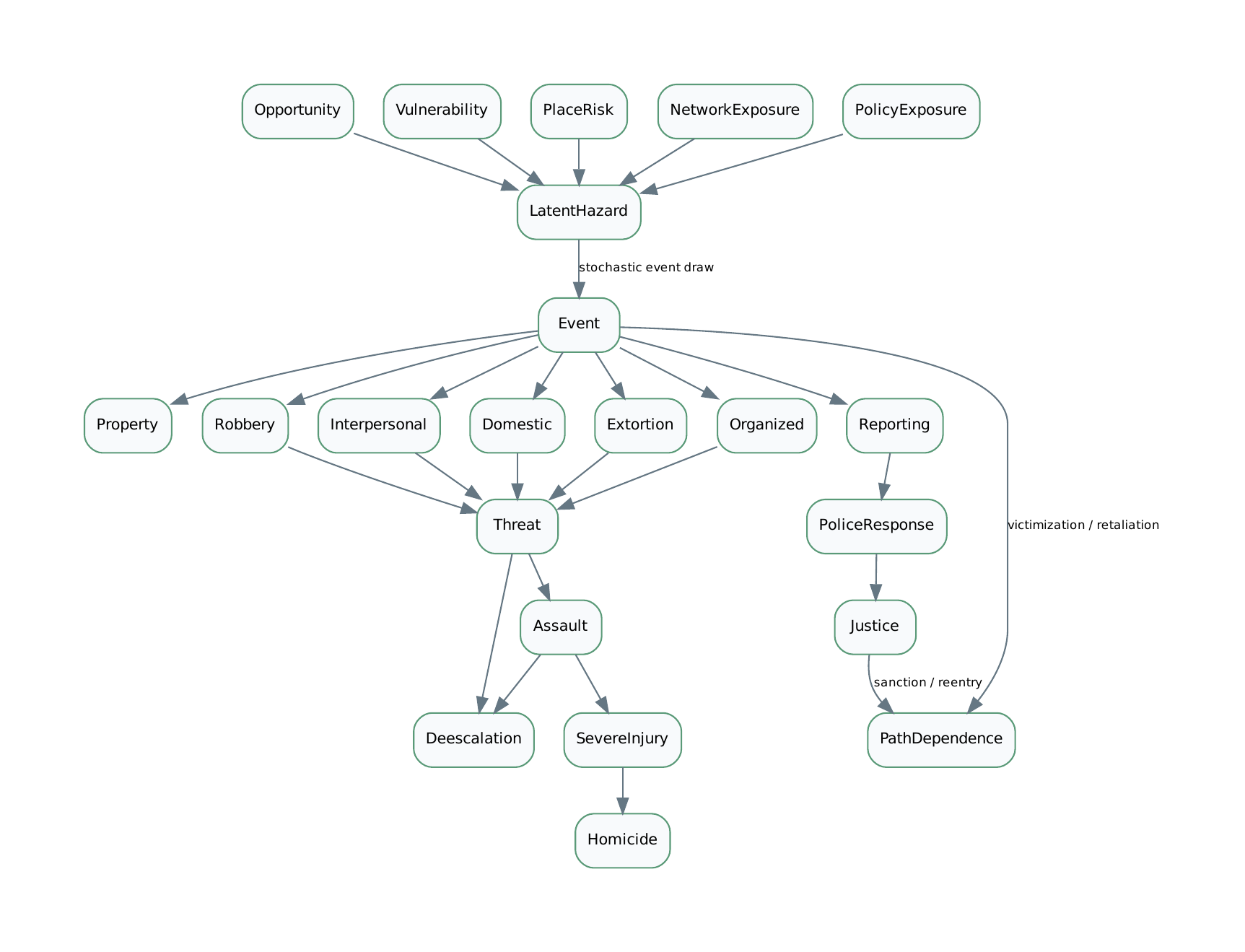}
\caption{Crime-event and violence-escalation logic. Homicide is reachable only through a severe-violence branch.}\label{fig:eventflow}
\end{figure}

Severe violence is modeled as an escalation process conditional on a violent event rather than as an independent Bernoulli draw. An encounter can de-escalate or progress to threat, assault, severe injury, and in rare cases homicide. Guardianship, violence interruption, investigation, weapon/organization context, and institutional response alter escalation probabilities. This architecture prevents the model from generating homicide simply because a demographic profile has a high baseline coefficient. Figure~\ref{fig:eventflow} summarizes the event logic.

Reporting is endogenous. A crime event is formally reported with a probability that depends on institutional trust, procedural-justice exposure, witness protection, and an empirically calibrated reporting shift. This is analytically important because a policy can improve reporting even if true victimization does not change. Consequently, more police records after a legitimacy reform are not automatically interpreted as more crime. The model reports latent simulated victimization and simulated formal reporting separately.

Policy modules are composable. Hot-spots policing reduces place risk primarily in concentrated locations but creates a displacement channel into untreated locations. Problem-oriented policing operates as a diagnosis-and-response multiplier and is not equated with simple patrol intensity. Focused deterrence affects robbery, interpersonal violence, extortion, and organized violence among high-harm networks more than low-level property crime. Procedural justice mainly changes trust and reporting, with only a deliberately small direct crime channel. Community social control acts on local compliance, extortion, conflict, and community information but can be degraded by capture. Guardabosques-type incentives combine community monitoring with an economic-substitution channel. Youth prevention, adult CBT, employment, and rehabilitation reduce dynamic latent risk rather than creating an instantaneous national crime shock. Lighting and alcohol regulation operate through place opportunity and situational violence. Investigation, witness protection, and financial-network disruption concentrate on severe and organized channels. Oversight primarily improves institutional legitimacy and reduces the risk that concentrated enforcement is implemented without accountability.

The intervention priors are deliberately conservative. Strong meta-analytic findings are not copied directly as national average treatment effects. For example, the approximate evidence for hot spots or focused deterrence concerns targeted high-risk units, not every person in Bolivia. The model therefore interprets external evidence as a treated-channel reduction before reach, capacity, fidelity, and diminishing returns. Community control and Guardabosques-type priors are even more strongly shrunk at the national level because the most persuasive evidence concerns conflict reduction, compliance, conditional homicide effects, illicit-crop control, or local governance in specific territories rather than universal reductions in all crime \parencite{farthingkohl2012,grisaffiledebur2016,brewer2021,farthinggrisaffi2024,idbguardabosques,dnpguardabosques}. This modeling decision explains why those policies produce relatively small aggregate victimization changes in the national-style development scenario even though they may be highly relevant in targeted contexts.

Capacity and resource constraints prevent unrealistic universal treatment. The code includes capacity and budget ledgers that cannot allocate more treatment than configured. The included scenario cost index is not a Boliviano budget estimate; it is a relative resource-intensity metric derived from lever intensity and agent-time exposure. It is used for Pareto comparison, not for fiscal appropriation. A future version with municipal and national unit costs could replace this index with monetary cost-effectiveness.

Displacement is explicit. When hot-spots or focused enforcement suppresses opportunity in targeted locations or networks, a fraction of the removed risk can reappear in less targeted places. The structural displacement fraction is not identified by the household surveys and is therefore drawn from a registered uncertainty interval. Similarly, capture is drawn as a structural uncertainty and weakens community-control mechanisms. Criminal adaptation increases latent risk after repeated exposure to enforcement and prior offending. These pathways make it possible for an intervention to look favorable on its immediate target while performing less well on the system-wide outcome.

\section{Ensemble emulation, global sensitivity, and Bayesian calibration}

Directly calibrating a stochastic ABM by running the full simulation inside every MCMC proposal is computationally expensive. The inference architecture therefore separates simulator design from emulator-based inference. A Latin-hypercube ensemble spans uncertain baseline parameters. The development calibration design varies the baseline crime-hazard scale, reporting shift, and trust drift. Each design point is simulated using a fixed, logged seed scheme. Outputs include victimization, reporting, trust, severe violence, and other internal metrics. The resulting design becomes the training set for an adaptive surrogate ensemble.

The first emulator is an automatic-relevance-determination Gaussian process with a Matérn kernel and predictive variance \parencite{rasmussen2006gp}. The second is a Bayesian-bootstrap tree ensemble. It uses bootstrapped extremely randomized trees and posterior-like bootstrap prediction draws; it is deliberately labeled \emph{BayesianTreeEnsemble} rather than BART because the execution environment did not contain an exact BART implementation. This naming constraint is part of the reproducibility tests and prevents a deterministic or approximate tree model from being presented as a different Bayesian method. The conceptual role is nevertheless similar to the motivation for Bayesian additive regression trees: provide a flexible nonlinear complement to the smooth Gaussian process \parencite{chipman2010bart}.

The two emulators are combined using weights derived from held-out predictive root-mean-square error. The stack retains predictive variance from within-model uncertainty and disagreement between models. Failed or non-finite emulators are not eligible for downstream sensitivity or calibration. The included tests recover a known nonlinear synthetic function with a low prediction error before the emulator is used on the ABM ensemble. Convergence warnings from Gaussian-process hyperparameter optimization are retained in execution logs rather than suppressed; they indicate that some length scales are weakly identified, which is consistent with the later sensitivity result that only a subset of policy channels materially drives the emulated response.

Adaptive enrichment is performed after the initial design. A provisional GP--tree stack scores a larger candidate design by predictive uncertainty and cross-model disagreement, and the highest-value candidate points are returned to the ABM before the final stack is fit. For the inner MCMC loop, the final nonlinear stack is then compiled into a quadratic ridge projection over 360 stack predictions. This compilation is a computational device rather than a new statistical model: held-out diagnostics compare the compiled projection against the final stack, with RMSEs of 0.00157 for victimization, 0.00159 for reporting, and 0.00031 for trust in the included development run. Calibration therefore targets the adaptive GP--tree predictive surface while avoiding a full stochastic ABM or tree-ensemble evaluation inside every MCMC proposal.

Global sensitivity uses variance-based Sobol indices and an active-subspace calculation \parencite{saltelli2010sobol,constantine2015active}. The Sobol analysis operates on the policy-response emulator, not directly on a small number of raw simulations. Five higher-level channels are varied in the included development design: hot-spots/place-focused intensity, community-control intensity, social-prevention intensity, lighting/environmental intensity, and investigation intensity. First-order and total-order indices are reported separately. Monte Carlo noise can make first-order estimates near zero slightly unstable, so substantive interpretation emphasizes the combination of total-order indices, emulator diagnostics, and active-subspace scores rather than a single negative or very small first-order value.

\begin{table}[H]\centering\small\caption{Global sensitivity and active-subspace diagnostics}\begin{tabular}{lrrr}\toprule Parameter & Sobol $S_1$ & Sobol $S_T$ & Activity score\\\midrule
hotspots & 1.000 & 0.984 & 0.00205\\
community control & -0.296 & 0.034 & 0.00001\\
social & -0.116 & 0.021 & 0.00056\\
lighting & 0.042 & 0.067 & 0.00001\\
investigation & -0.059 & 0.014 & 0.00001\\
\bottomrule\end{tabular}\end{table}

\begin{figure}[H]\centering
\includegraphics[width=.82\textwidth]{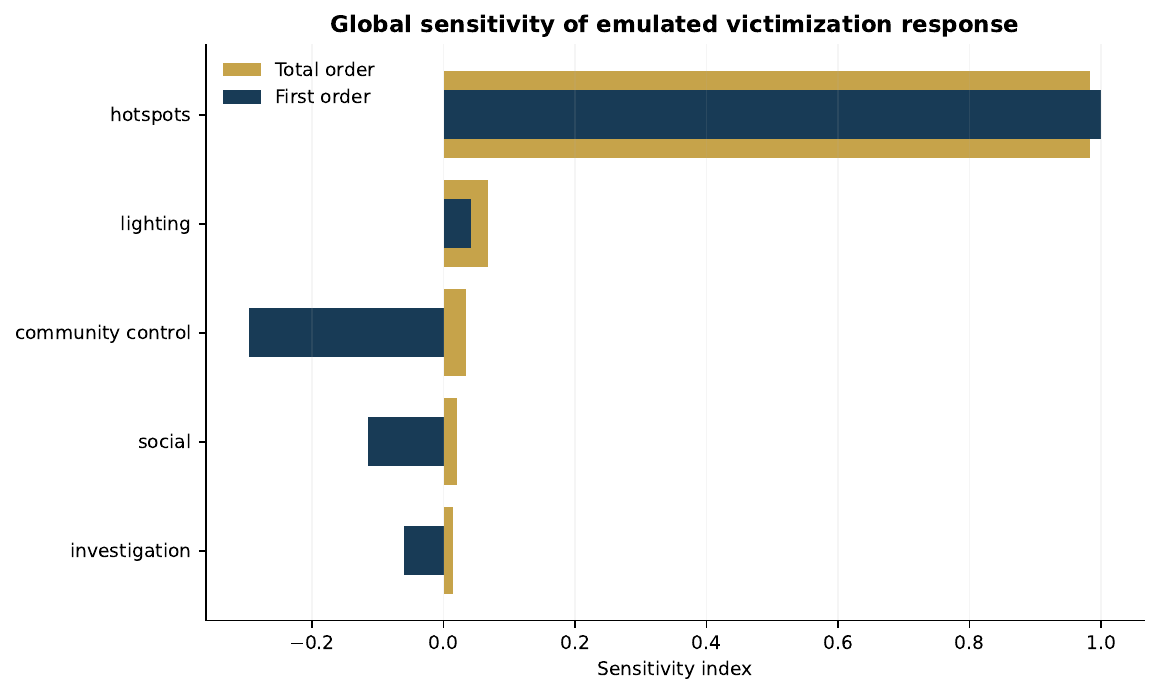}
\caption{Sobol sensitivity indices for the emulated victimization response. Indices are conditional on the development design and should not be interpreted as empirical causal shares.}\label{fig:sobol}
\end{figure}

\begin{figure}[H]\centering
\includegraphics[width=.76\textwidth]{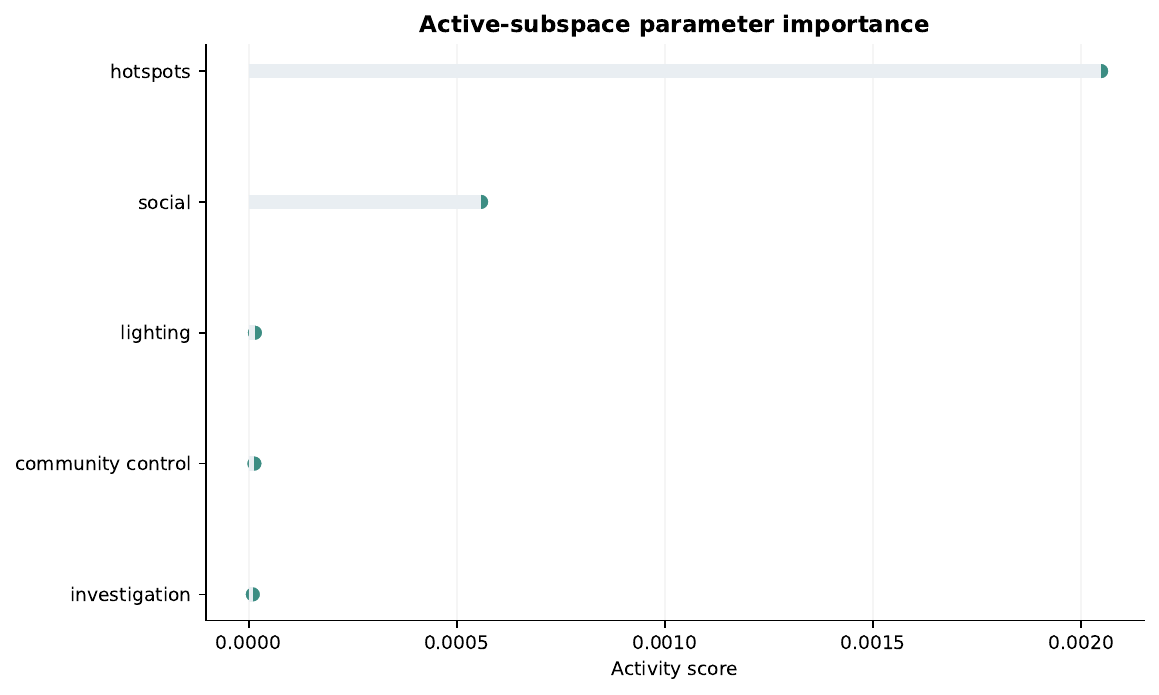}
\caption{Active-subspace activity scores. The first active direction is dominated by the place-focused policy channel.}\label{fig:active}
\end{figure}

The development sensitivity analysis identifies the place-focused hot-spots channel as the dominant driver of the emulated short-run victimization response. Its estimated total-order index is about 0.984. Lighting has a materially smaller total-order contribution, and community-control, social, and investigation channels are smaller in this particular national-average, 12-month emulated outcome. This does not mean that those channels are unimportant for homicide, organized crime, legitimacy, long-run recurrence, or local settings. It means that under the model's parameterization and the outcome selected for this sensitivity exercise, short-run overall victimization responds most strongly to place concentration. The active-subspace direction points to the same dominant dimension, as shown in Figures~\ref{fig:sobol} and \ref{fig:active}.

Baseline calibration is conducted with a parallel-tempering style ensemble MCMC using multiple temperatures and random-walk ensemble proposals, with periodic covariance adaptation used as dynamic preconditioning \parencite{earl2005parallel}. The likelihood compares compiled-emulator predictions for victimization, reporting, and trust against the weighted EH targets while allowing explicit model discrepancy. Three independent tempered calibrations are run with discrepancy multipliers 0.80, 1.00, and 1.25. Each contributes 570 post-burn-in cold-chain draws. Their proposal acceptance rates are 0.477, 0.545, and 0.600, with adjacent-temperature swap rates 0.343, 0.384, and 0.376. A predictive meta-ensemble weights the three calibrations by out-of-sample target discrepancy, producing weights 0.498, 0.329, and 0.173 and 1710 stacked posterior draws. This is not claimed to be a definitive posterior for crime in Bolivia; it is a model-conditional calibrated distribution over three structural baseline parameters.

\begin{table}[H]\centering\small\caption{Predictive meta-ensemble posterior summary after survey-weighted population calibration}\begin{tabular}{lrrr}\toprule Parameter & Median & 5th percentile & 95th percentile\\\midrule
crime scale & 1.330 & 0.985 & 1.728\\
reporting shift & -0.162 & -0.179 & -0.114\\
trust shift & 0.008 & -0.079 & 0.084\\
\bottomrule\end{tabular}\end{table}

\begin{figure}[H]\centering
\includegraphics[width=.92\textwidth]{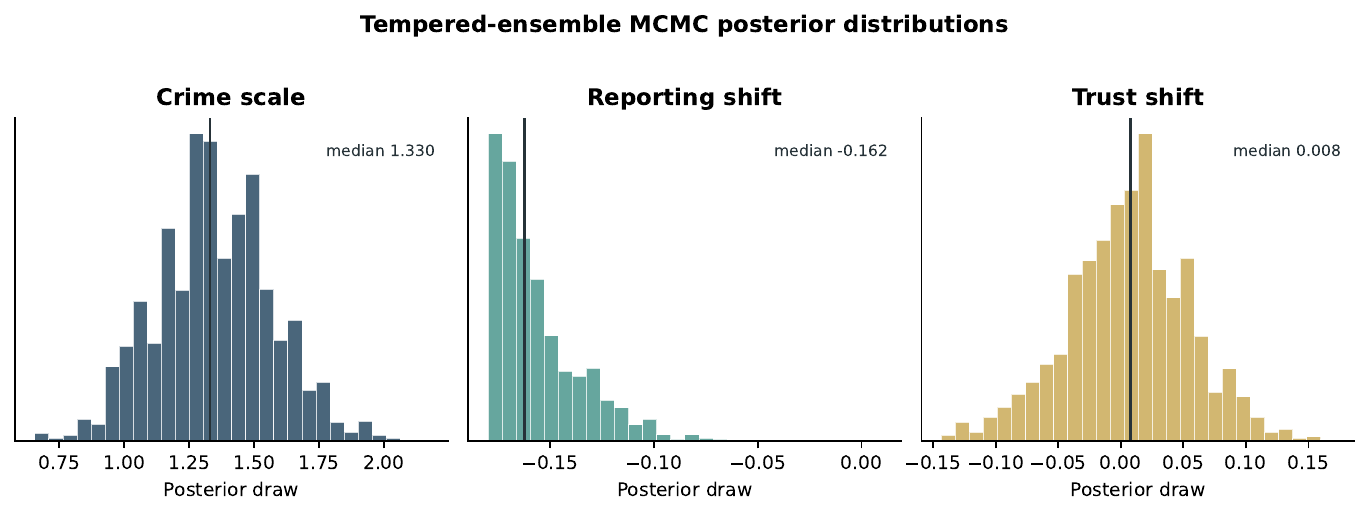}
\caption{Tempered-ensemble MCMC posterior distributions for the three survey-calibrated structural parameters.}\label{fig:post}
\end{figure}

The predictive meta-ensemble posterior median baseline crime-scale parameter is approximately 1.330, with a 5th--95th percentile interval of about 0.985--1.728. The reporting shift is negative, with a median of roughly -0.162, reflecting the low observed reporting rate relative to the uncalibrated structural reporting equation. The trust drift remains centered close to zero at approximately 0.008 and is less tightly identified. Figure~\ref{fig:post} shows the posterior distributions. The interpretation is deliberately structural: these parameters make the baseline ABM reproduce the observed summaries more closely; they are not regression coefficients estimated directly from respondents.

\begin{figure}[H]\centering
\includegraphics[width=.84\textwidth]{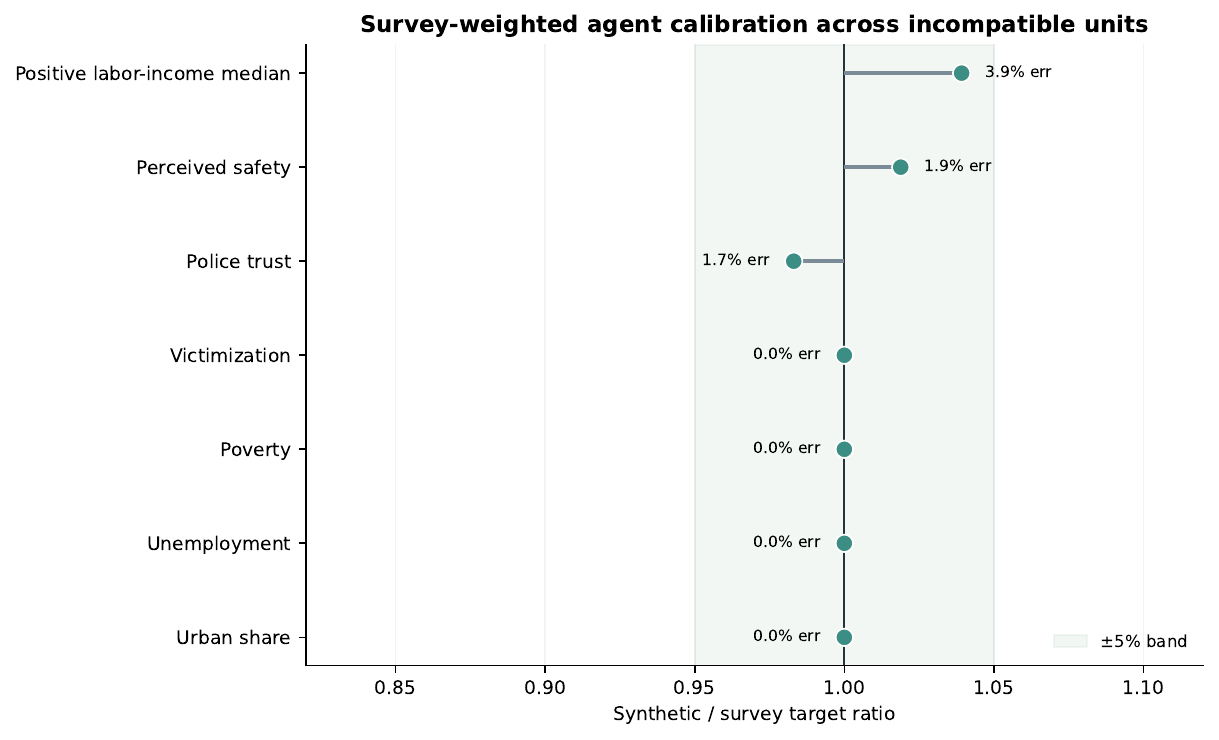}
\caption{Survey-weighted synthetic-agent calibration shown as synthetic-to-survey ratios so indicators with different units remain comparable.}\label{fig:popcal}
\end{figure}

\begin{figure}[H]\centering
\includegraphics[width=.84\textwidth]{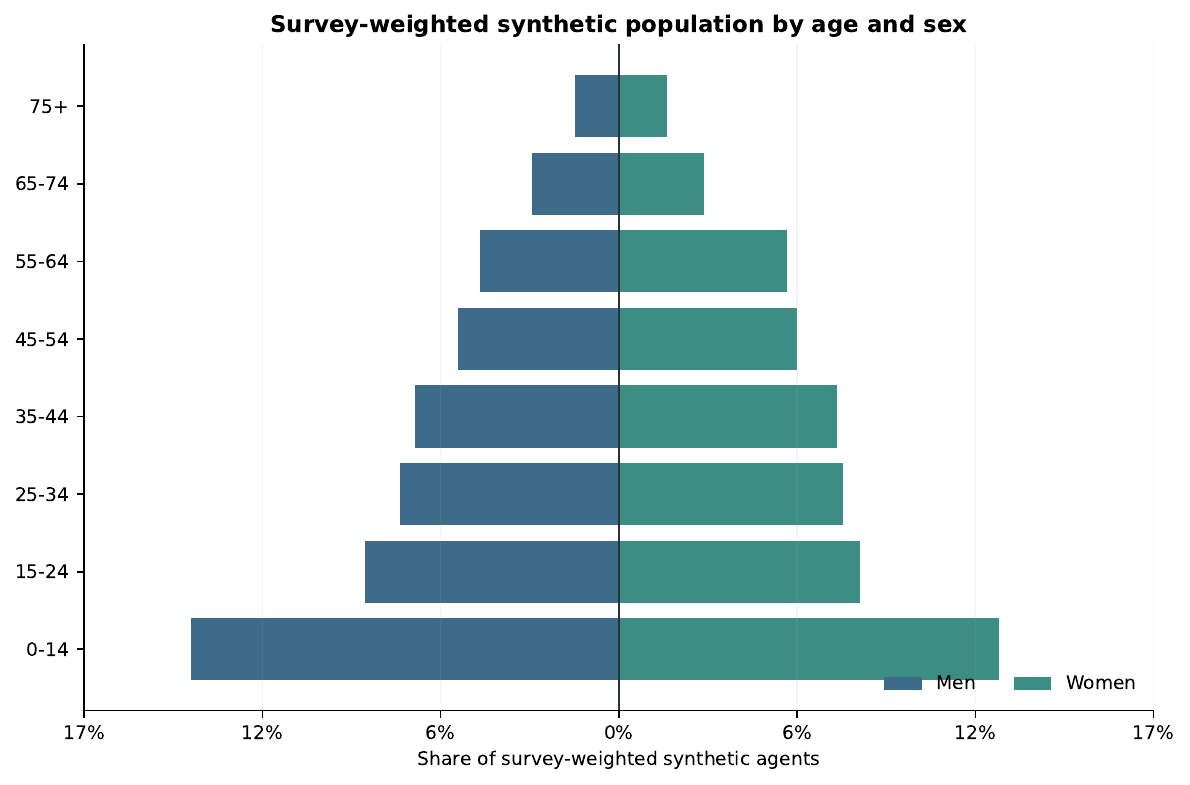}
\caption{Age-sex structure of the survey-weighted synthetic agents. Household/person selection uses EH expansion factors; labor and security traits are added from separately weighted ECE and EH donor pools.}\label{fig:agentpyramid}
\end{figure}

Synthetic-population validation is performed independently of crime calibration. Agents are now created directly from the survey designs rather than from independent draws around aggregate marginals. EH-2025 provides the demographic, household, poverty, food-security, and household-income backbone: households are resampled with probability proportional to the official expansion factor so that within-household dependence is retained for most synthetic households. ECE 4T-2025 provides quarterly labor-market donors matched by department, urban/rural area, sex, and age band and sampled with \texttt{fact\_trim\_act}. The EH discrimination/security module provides victimization, trust, and perceived-safety donors matched on the same observable cells and sampled with \texttt{PONDERAD}. Final synthetic-agent analysis weights are raked to the survey targets for urban residence, poverty, working-age unemployment, and adult victimization. The ABM then uses those weights when aggregating victimization, severe violence, reporting, trust, and policy cost. This creates survey-weighted agents while avoiding any claim that individual respondents are being reconstructed. Figure~\ref{fig:popcal} reports unit-free calibration ratios, which prevents labor income from visually overwhelming probability and score targets. Figure~\ref{fig:agentpyramid} shows the resulting weighted age-sex structure. The public replication package contains synthetic outputs and source design weights but no original respondent identifiers or microrecords.

\begin{figure}[H]\centering
\includegraphics[height=.70\textheight,keepaspectratio]{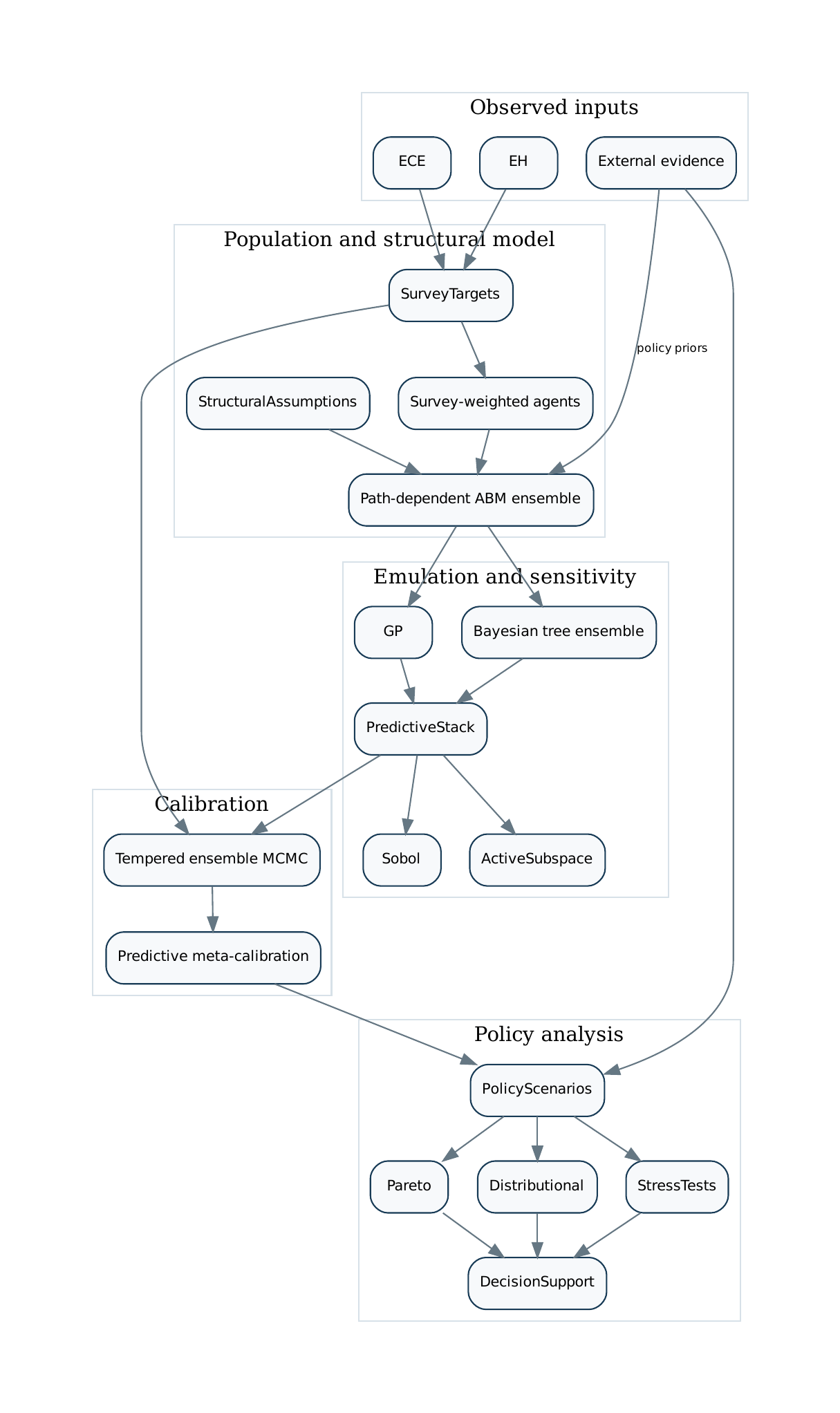}
\caption{Inference and decision-analysis pipeline.}\label{fig:infer}
\end{figure}

Figure~\ref{fig:infer} displays the full computational chain. Survey targets inform synthetic population generation and likelihood calibration. External evidence informs policy priors. Structural assumptions enter the ABM ensemble. The Gaussian-process and Bayesian-tree surrogates support sensitivity and MCMC. Posterior draws are then combined with intervention-prior and structural draws in policy scenarios. Multi-objective results feed Pareto, distributional, and stress-test analyses. This separation is what allows the package to state which part of a result comes from observed Bolivian data and which part comes from assumptions or transported evidence.

\section{Policy packages, simulated results, and trade-offs}

Seventeen scenario families are evaluated with common random numbers across scenarios within each posterior/structural replicate. Pairing is important. Without paired seeds, random differences in places, event draws, or severity could be mistaken for policy effects. Each replicate draws a calibrated baseline parameter vector, policy-effect channel values from conservative intervention priors, and structural values for violence severity, criminal adaptation, institutional capture, and displacement. The same replicate-level parameter draw and stochastic seed are then reused across all policy packages. The resulting difference from baseline therefore removes much of the Monte Carlo noise that would otherwise contaminate scenario comparisons.

The baseline scenario contains no policy lever and functions only as a paired comparator. Focused policing combines hot-spots and problem-oriented policing. The focused-deterrence scenario adds witness protection. Community control combines local monitoring with procedural legitimacy. The Guardabosques scenario adds conditional collective incentives and an employment/economic-substitution channel. Social prevention combines youth programs, adult CBT, employment, and rehabilitation. The environmental scenario combines lighting with alcohol regulation. Justice legitimacy emphasizes procedural justice, oversight, and witness protection. Investigation emphasizes homicide/investigative capacity and witness protection. Organized-crime disruption combines financial/network investigation, homicide investigation, and focused deterrence. Balanced, enforcement-heavy, social-heavy, community-heavy, place-based, rights-first, and full-integrated portfolios explore alternative combinations and intensities.

The full scenario table is reported below. The percentages are simulated paired changes relative to baseline. They must not be interpreted as empirical effect sizes for Bolivia. In particular, the cost index is dimensionless, and Pareto status depends on the selected objectives and the finite scenario set.

\begin{table}[H]\centering\scriptsize\caption{Development-run scenario results using survey-weighted synthetic agents. All effects are model-conditional scenarios, not causal estimates for Bolivia.}\resizebox{\textwidth}{!}{\begin{tabular}{lrrrrrr}\toprule Scenario & Victim. \%$\Delta$ & Severe \%$\Delta$ & Trust \%$\Delta$ & Cost index & Displacement & Pareto\\\midrule
full integrated & 17.9 & 19.3 & 13.5 & 295.9 & 0.176 & Yes\\
place based & 17.4 & 13.1 & 0.5 & 106.1 & 0.163 & Yes\\
enforcement heavy & 14.9 & 20.0 & 0.8 & 177.9 & 0.312 & Yes\\
balanced & 14.1 & 14.4 & 5.3 & 141.5 & 0.156 & Yes\\
focused policing & 13.4 & 11.7 & 0.4 & 59.7 & 0.176 & Yes\\
social heavy & 4.9 & 5.8 & 0.3 & 135.2 & 0.000 & Yes\\
environmental & 4.4 & 3.3 & 0.1 & 44.9 & 0.000 & Yes\\
social prevention & 3.6 & 4.6 & 0.2 & 90.8 & 0.000 & Yes\\
focused deterrence & 3.5 & 6.5 & 3.4 & 63.6 & 0.088 & Yes\\
organized crime & 3.3 & 7.3 & 0.3 & 127.5 & 0.054 & Yes\\
community heavy & 2.5 & 2.7 & 6.9 & 77.9 & 0.000 & No\\
community control & 1.6 & 1.5 & 4.8 & 31.4 & 0.000 & Yes\\
guardabosques & 1.6 & 1.9 & 0.1 & 69.2 & 0.000 & No\\
rights first & 1.0 & 1.0 & 23.1 & 80.3 & 0.000 & No\\
investigation & 0.6 & 3.9 & 5.3 & 87.1 & 0.000 & Yes\\
baseline & 0.0 & 0.0 & 0.0 & 0.0 & 0.000 & Yes\\
justice legitimacy & 0.0 & 0.0 & 19.1 & 50.7 & 0.000 & No\\
\bottomrule\end{tabular}}\end{table}

The full integrated package produces the largest mean paired reduction in overall victimization among the evaluated portfolios, approximately 17.9\%, with a paired 5th--95th percentile interval of roughly 15.1--20.9\%. Its paired mean reduction in severe violence is about 19.3\%, with a corresponding interval of approximately 16.4--21.8\%. The ratio of aggregate scenario means reported in the scenario table is 17.9\% for victimization and 19.3\% for severe violence; the small difference from the paired mean reflects nonlinear averaging across replicates. It also increases simulated trust by roughly 13.5\%, but it has by far the largest resource-cost index among the scenarios and retains a non-trivial displacement index. The result therefore does not imply that the full package is the preferred real-world policy. It illustrates how a broad portfolio can perform when many channels are activated simultaneously and their effects are subjected to diminishing returns and uncertainty.

The place-based package combines hot-spots, problem-oriented policing, lighting, and alcohol regulation. Its paired victimization reduction averages about 17.5\% (15.1--21.9\% at the 5th--95th percentiles), close to the integrated package but at a much lower resource-cost index. The paired severe-violence reduction is smaller, roughly 13.6\%. This difference is structurally coherent: place opportunity is strongly related to common victimization, whereas organized and severe violence also depends on network, investigation, and escalation channels. Focused policing alone reduces simulated victimization by about 13.4\% and severe violence by 11.7\%, with a paired victimization interval around 9.7--17.0\%. The result is consistent with the global sensitivity analysis, but it is conditional on the model's transport of hot-spots and POP evidence.

The enforcement-heavy portfolio produces a paired victimization reduction of about 15.0\% and a paired severe-violence reduction of roughly 19.5\%, but has the largest displacement index in the scenario set, approximately 0.312. This is an important example of why ranking policies solely by violence reduction can be misleading. A strategy that concentrates enforcement can produce a strong immediate reduction in severe events while increasing the risk that activity shifts to other places or networks. The model does not assert that this displacement magnitude is observed in Bolivia; it shows how a policy can cease to dominate once a plausible displacement mechanism is included.

The balanced package produces paired reductions of approximately 14.1\% in victimization and 14.1\% in severe violence, while the aggregate scenario means imply a trust increase of about 5.3\%. Its cost is higher than focused policing but substantially below the full integrated package. In the finite scenario set it remains Pareto non-dominated under the selected dimensions. The rights-first package generates only about 1.0\% reduction in victimization and 1.0\% in severe violence but raises trust by roughly 23.1\%. Justice legitimacy similarly produces essentially no direct victimization reduction in the implementation but increases trust by about 19.1\%. This is intentional: procedural justice is not allowed to masquerade as a large direct crime treatment. Its principal modeled outcomes are legitimacy and reporting.

Community-control and Guardabosques-style scenarios yield aggregate-mean national-style victimization reductions of roughly 1.6\% and 1.6\%, respectively, in the development simulation. The low aggregate effect is a modeling choice grounded in external-validity caution, not a finding that these programs are ineffective. The strongest evidence for Bolivian coca social control concerns reductions in confrontation, improved local governance, livelihood security, and compliance in specific producer territories \parencite{farthingkohl2012,grisaffiledebur2016,brewer2021,farthinggrisaffi2024}. The Colombian Guardabosques evidence combines local monitoring, conditional incentives, alternative development, and changes in homicide or illicit crops in specific program phases \parencite{idbguardabosques,dnpguardabosques}. The present national-style population simulation therefore does not extrapolate those effects to every type of crime or every municipality. A future targeted Chapare, Yungas, border, or high-illicit-market microsimulation could assign much higher reach to those mechanisms and answer a different question.

Social-heavy and social-prevention scenarios generate smaller short-run aggregate changes, approximately 4.9\% and 3.6\% reductions in victimization, respectively. Again, this does not imply that developmental prevention is weak. The simulated horizon is two years and many social interventions operate through longer-term offending trajectories, education, employment, recidivism, and intergenerational mechanisms. A short-horizon ABM will mechanically favor interventions that affect immediate exposure unless long-run benefits are explicitly accumulated. The report therefore treats social prevention as a complementary component whose evaluation should include five- and ten-year horizons in production runs.

\begin{figure}[H]\centering
\includegraphics[width=.88\textwidth]{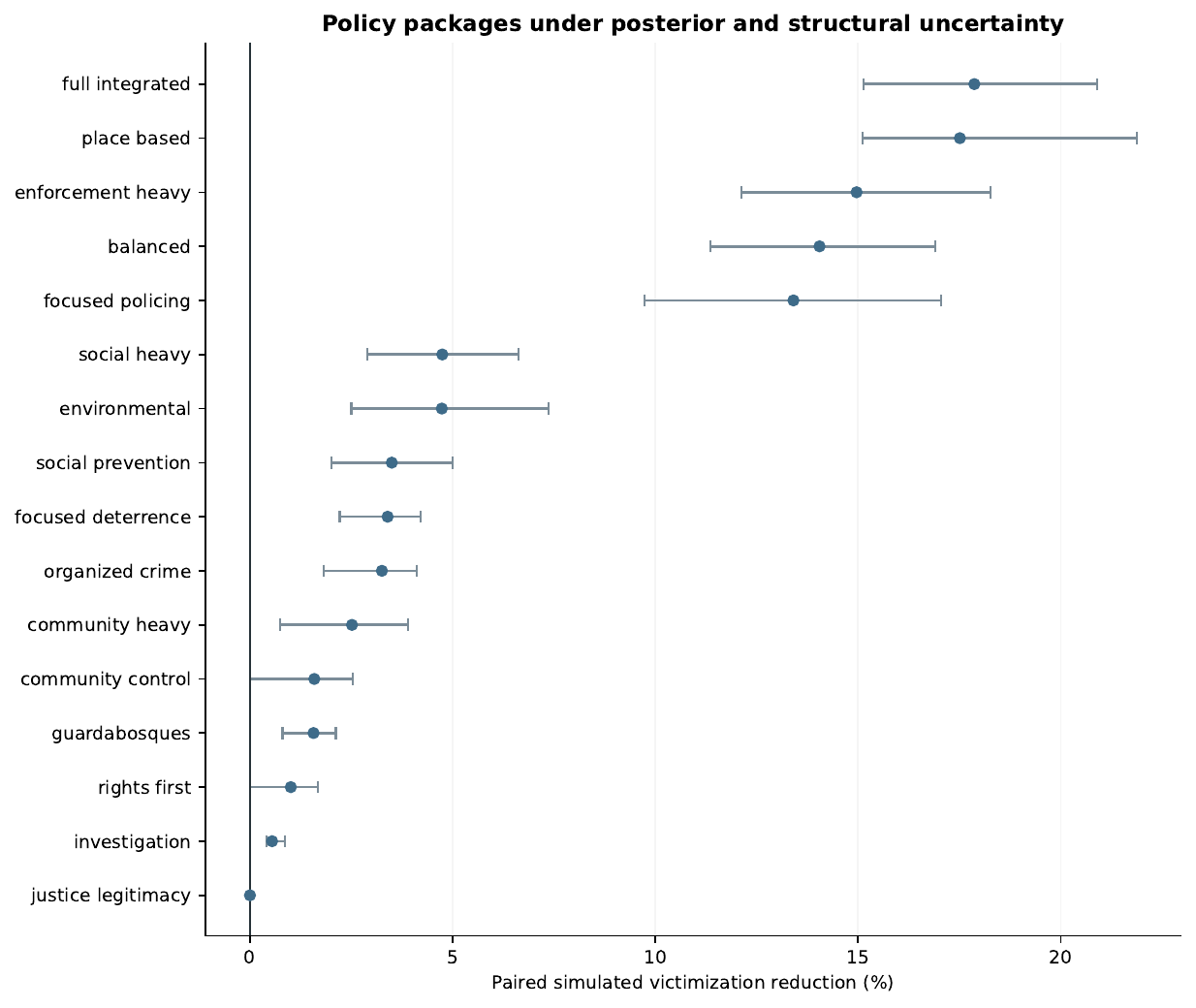}
\caption{Paired simulated victimization changes with 5th--95th percentile intervals across posterior, policy-prior, structural, and stochastic uncertainty. Values are model-conditional scenario effects.}\label{fig:uncertainty}
\end{figure}

Figure~\ref{fig:uncertainty} demonstrates why uncertainty matters. Several modest-effect scenarios have intervals close to zero in the limited development ensemble, while the place-based, focused-policing, balanced, enforcement-heavy, and full-integrated portfolios remain clearly separated from zero. The intervals do not include every conceivable model-form uncertainty; they propagate the uncertainties explicitly encoded in the development design. Structural alternatives that change network topology, weapon availability, municipal capacity, or organized-crime response could widen them substantially.

\begin{figure}[H]\centering
\includegraphics[width=.94\textwidth]{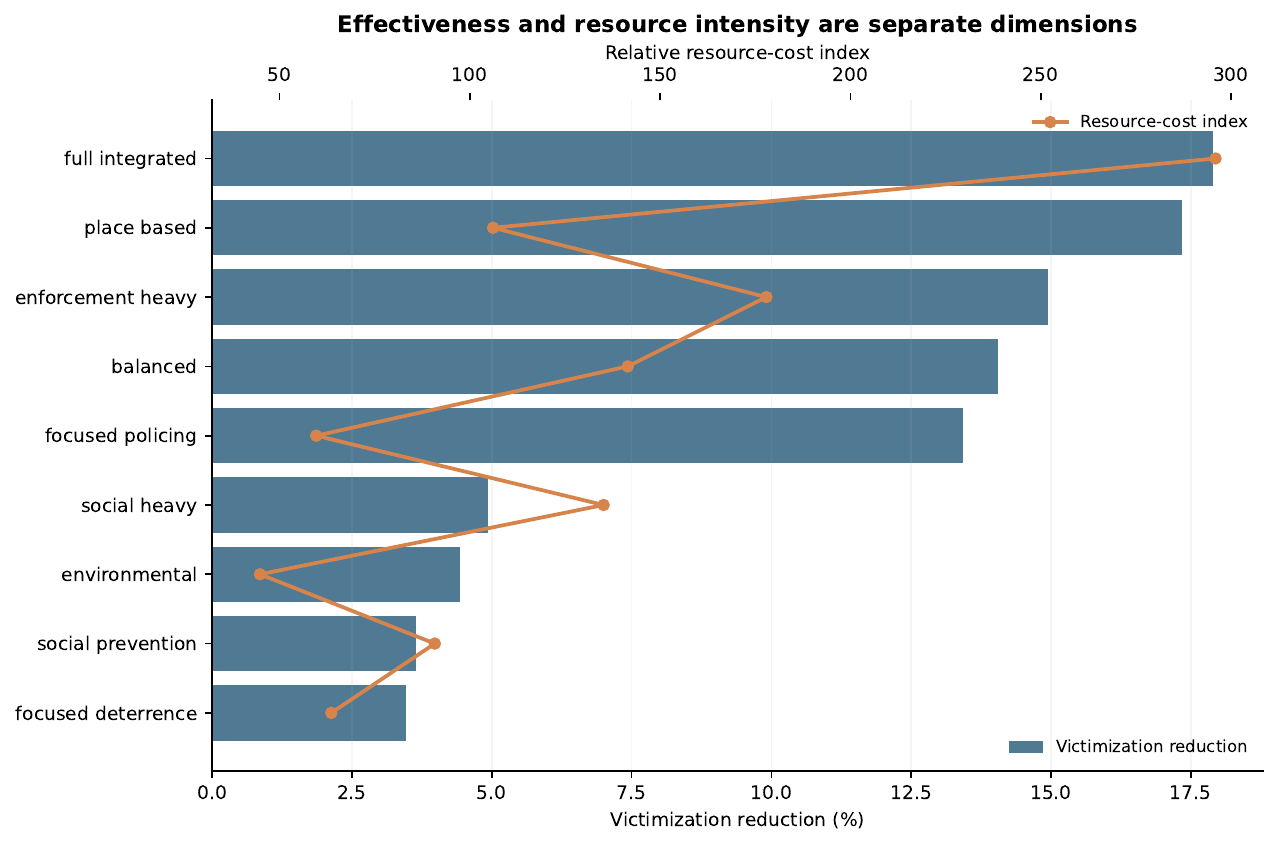}
\caption{Dual-axis comparison of simulated victimization reduction and the model's relative resource-cost index.}\label{fig:dual}
\end{figure}

Cost-effectiveness and impact magnitude are distinct. Figure~\ref{fig:dual} plots victimization reduction and the resource-cost index on separate axes. The full integrated package has the strongest reduction but the highest cost. Place-based and focused-policing packages provide much of the simulated victimization reduction at lower cost. Investigation and rights-oriented scenarios purchase different outputs: trust, reporting, organized-crime capacity, or procedural safeguards rather than the largest immediate decline in overall victimization.

\begin{figure}[H]\centering
\includegraphics[width=.92\textwidth]{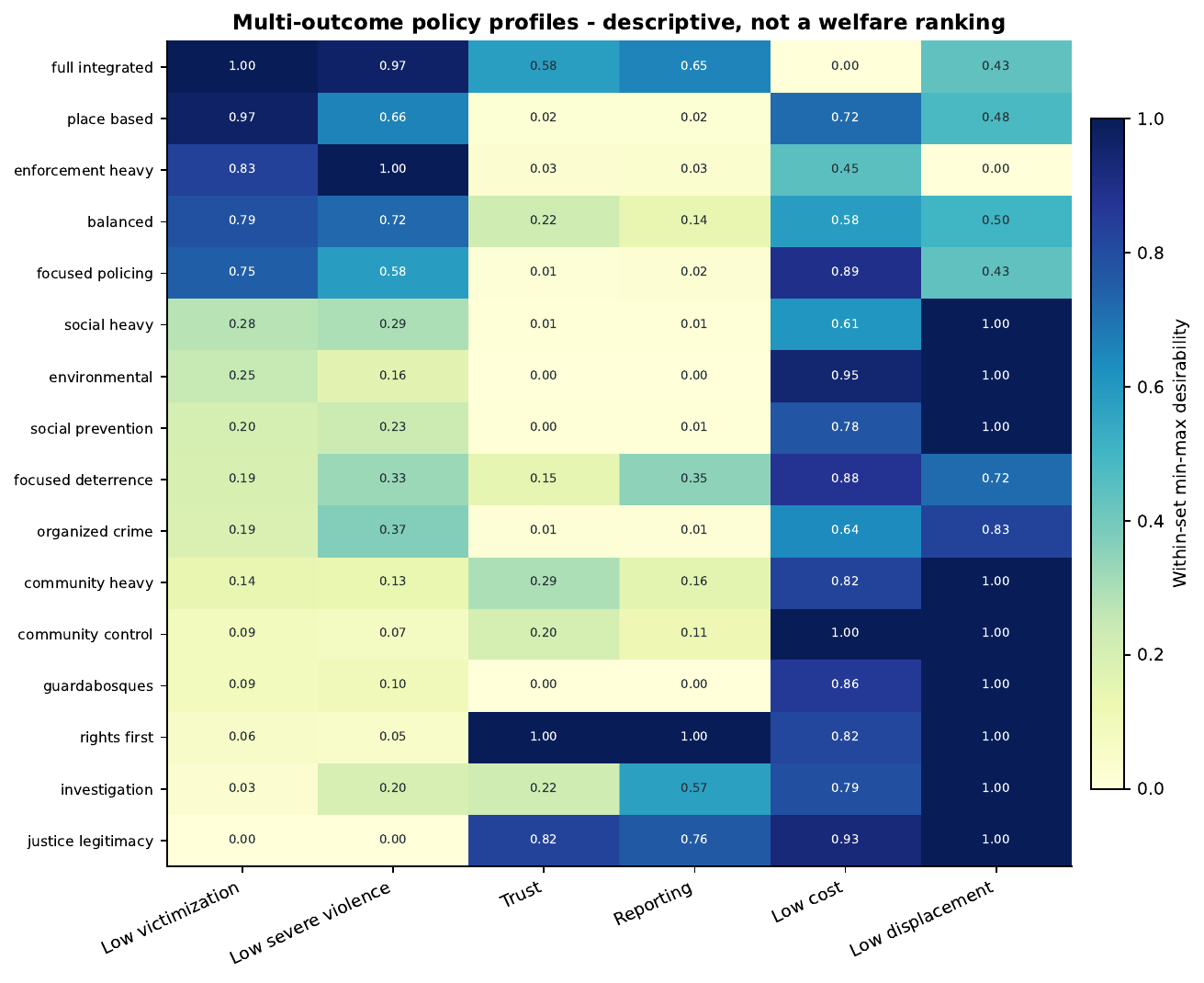}
\caption{Within-scenario-set min--max profiles. The display is a descriptive normalization, not a causal score or political ranking.}\label{fig:heat}
\end{figure}

A multi-outcome min--max display makes these differences clearer. The columns in Figure~\ref{fig:heat} normalize outcomes only within the scenario set and orient them so higher values correspond to lower victimization, lower severe violence, higher trust, higher reporting, lower cost, or lower displacement. It is not a social-welfare function and it is not a probability. The figure is used to prevent one outcome from silently determining the conclusion.

\begin{figure}[H]\centering
\includegraphics[width=.84\textwidth]{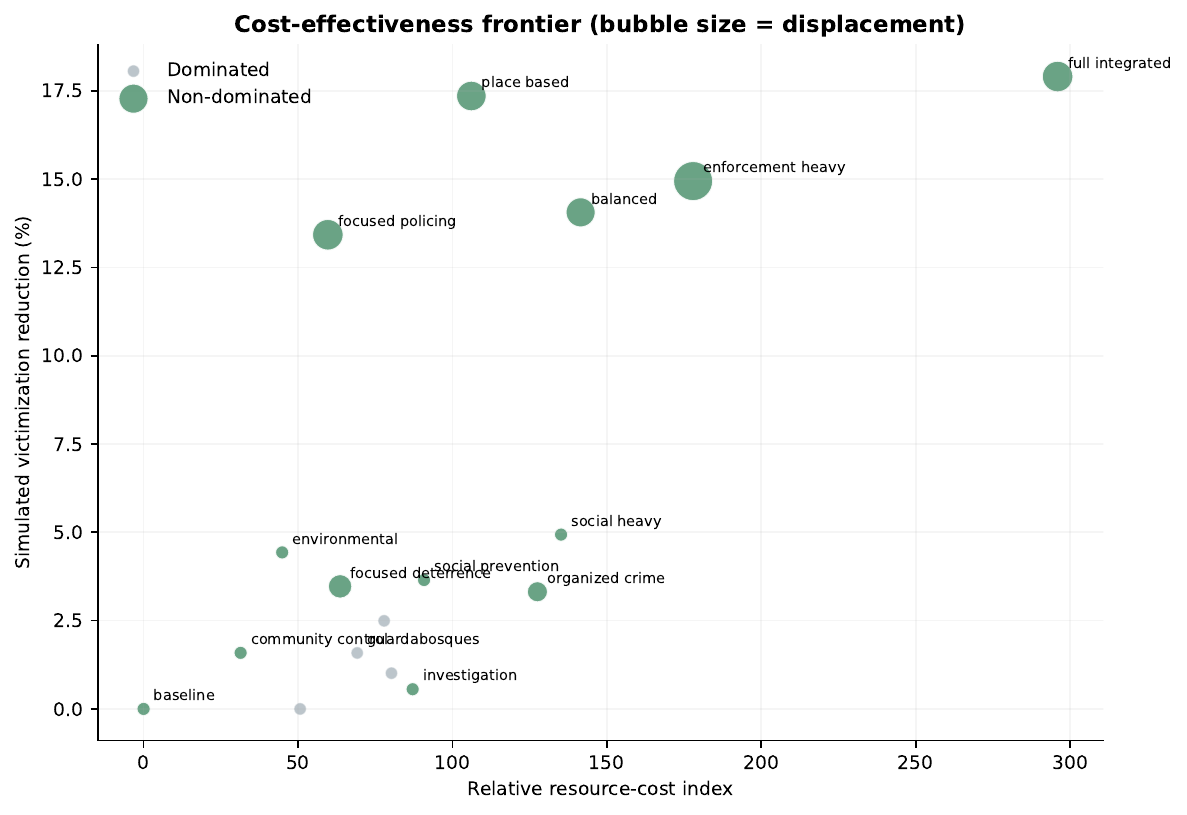}
\caption{Cost-victimization trade-off. Bubble size represents simulated displacement; green points are non-dominated in the selected multi-objective space.}\label{fig:pareto}
\end{figure}

The same principle motivates the Pareto frontier. A scenario is marked non-dominated if no other evaluated scenario is simultaneously no worse on victimization, severe violence, resource cost, and displacement and strictly better on at least one of them. Pareto status does not choose among non-dominated scenarios because that requires normative weights. Figure~\ref{fig:pareto} shows the cost-victimization plane, with displacement represented by bubble size. The baseline itself can remain Pareto non-dominated when zero cost is treated as an objective. This is mathematically correct and substantively useful: a policy analyst cannot claim that a costly intervention dominates doing nothing unless it improves every included objective enough to overcome its cost dimension.

Distributional evaluation is also necessary. The package conducts a paired stress test for the full integrated portfolio across poverty status, urban/rural residence, and age group. This is not a causal subgroup analysis from the EH. Synthetic groups are subsets of the survey-weighted fused agents, and the same policy architecture is simulated separately with common seeds and weighted aggregation. The results show mean reductions of roughly 16.7--17.4\% across the reported groups, with overlapping uncertainty ranges. The non-poor group has a modestly larger mean reduction than the poor group, and adults aged 30 or more have a modestly larger mean reduction than youth aged 15--29. These differences are small relative to the structural uncertainty and should be treated as diagnostics for further work rather than evidence of equitable real-world treatment effects.

\begin{table}[H]\centering\small\caption{Distributional stress test for the full integrated package using survey-weighted agents}\begin{tabular}{lrrrr}\toprule Group & Synthetic $n$ & Mean reduction & P05 & P95\\\midrule
Poor & 2287 & 16.7 & 16.4 & 17.2\\
Rural & 1482 & 16.7 & 16.2 & 17.4\\
Urban & 3518 & 17.0 & 16.5 & 17.3\\
Youth 15-29 & 1155 & 17.2 & 16.5 & 18.4\\
Age 30+ & 2526 & 17.3 & 16.3 & 18.0\\
Non-poor & 2713 & 17.4 & 17.1 & 17.8\\
\bottomrule\end{tabular}\end{table}

\begin{figure}[H]\centering
\includegraphics[width=.80\textwidth]{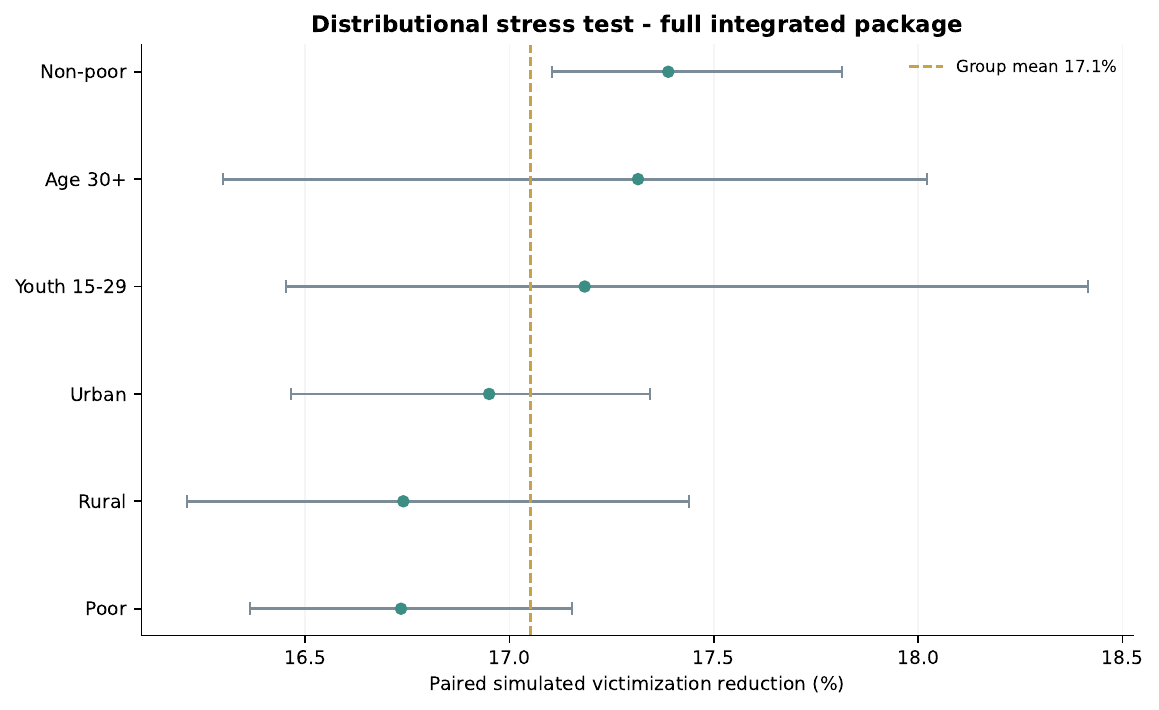}
\caption{Distributional stress test of the full integrated package in the synthetic population.}\label{fig:fair}
\end{figure}

\begin{figure}[H]\centering
\includegraphics[width=.78\textwidth]{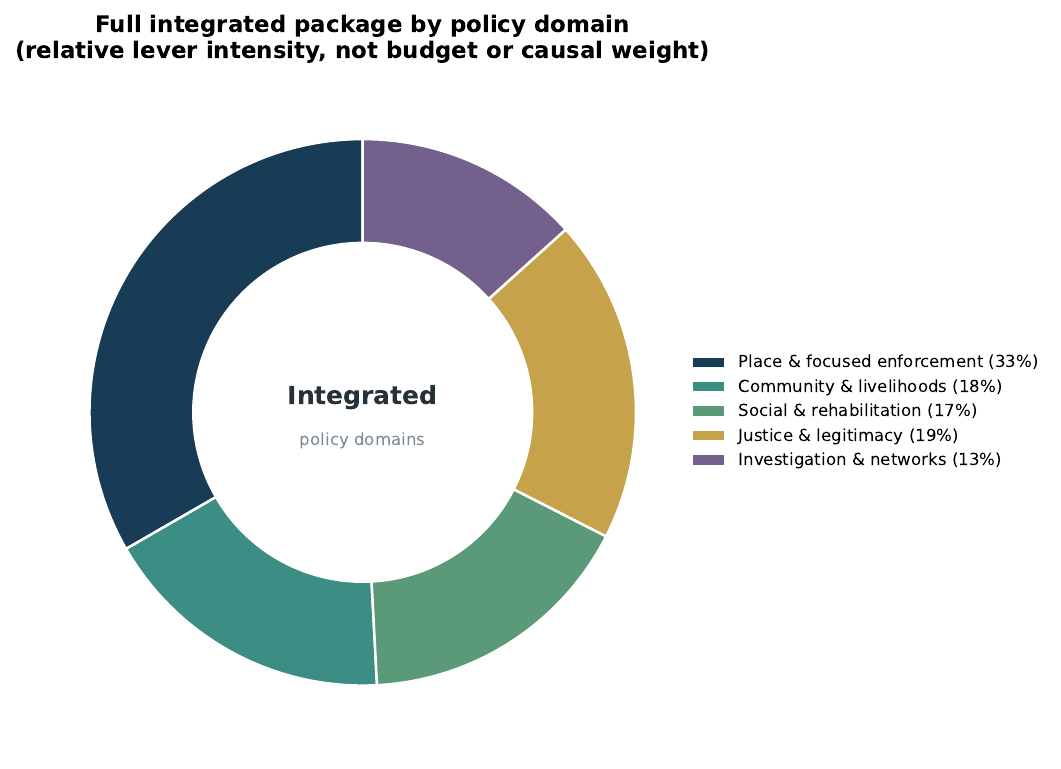}
\caption{Relative design intensity of the full integrated package aggregated into five policy domains. Percentages refer only to the sum of configured lever intensities, not budgets or causal weights.}\label{fig:donut}
\end{figure}

\begin{figure}[H]\centering
\includegraphics[width=.82\textwidth]{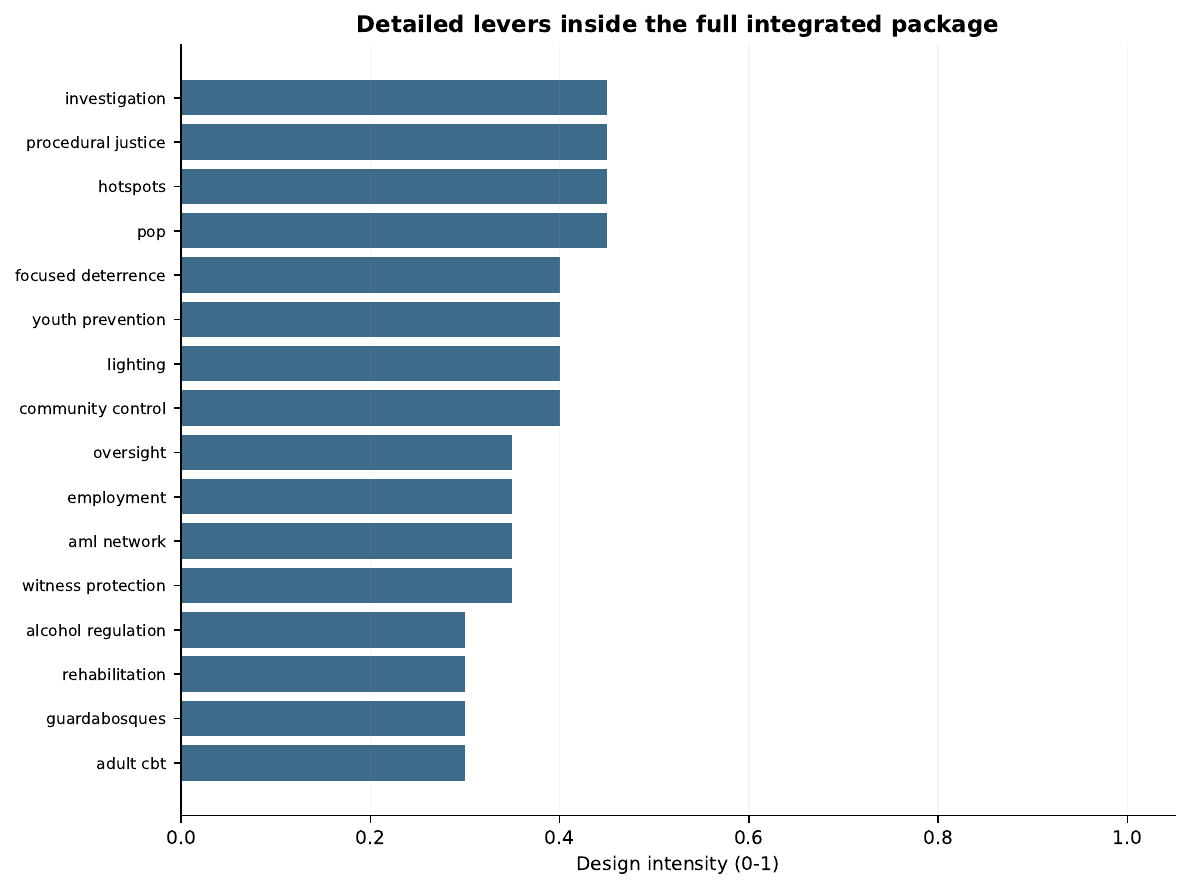}
\caption{Individual levers inside the full integrated package on their configured zero-to-one intensity scale.}\label{fig:levers}
\end{figure}

The full integrated package is itself a portfolio, not a single intervention. Figure~\ref{fig:donut} aggregates its lever intensities into five policy domains so that the circular display remains readable; Figure~\ref{fig:levers} reports every individual lever on a common zero-to-one scale. Neither display represents budget shares or estimated causal contributions. Their purpose is to show that the strongest aggregate scenario result does not come from one technology; it comes from simultaneous place, network, community, prevention, legitimacy, investigative, and rehabilitation channels, with diminishing returns and uncertainty.

\section{Decision rules, failure modes, and implementation}

\begin{figure}[H]\centering
\includegraphics[width=.98\textwidth]{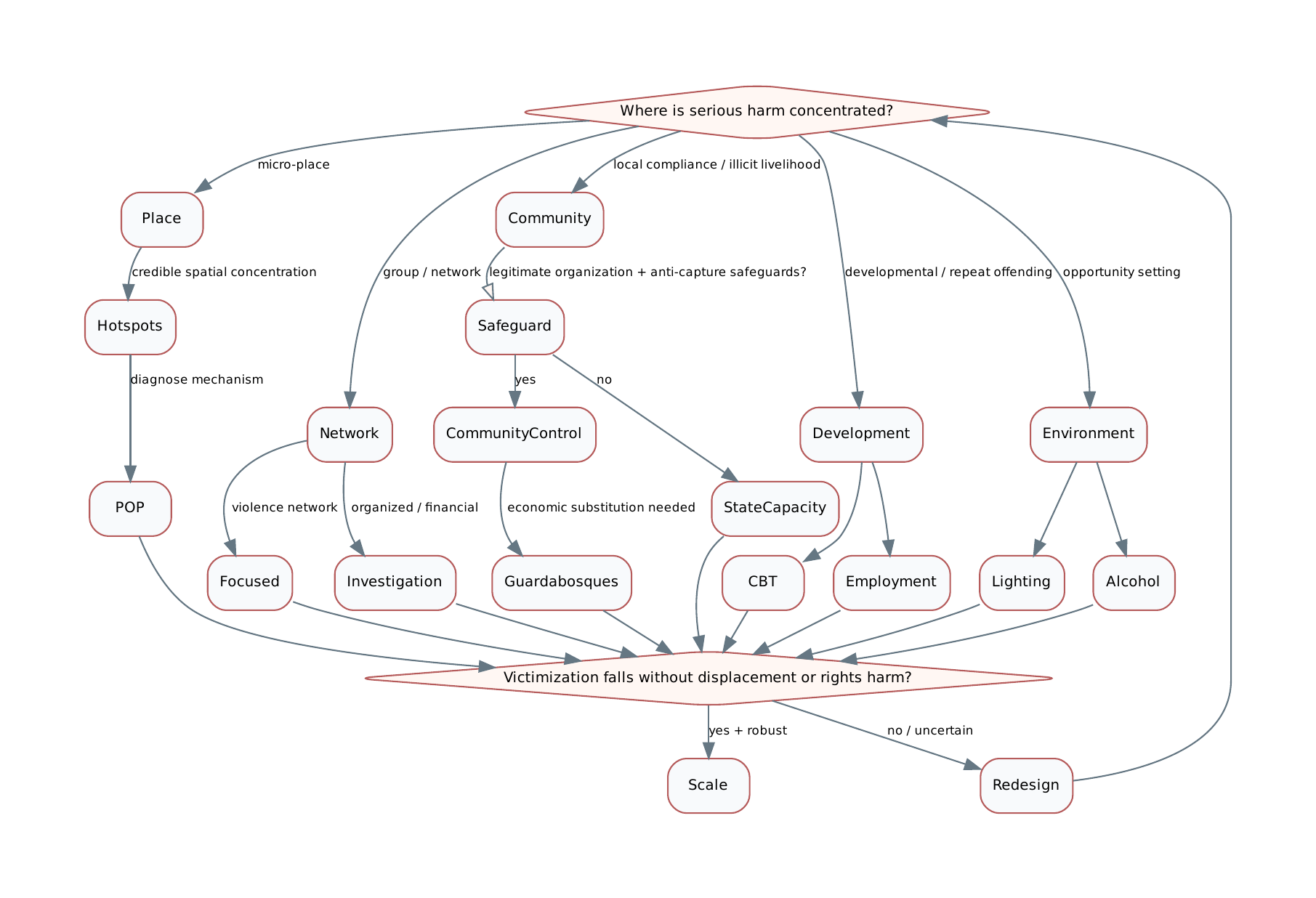}
\caption{Decision logic linking the dominant harm mechanism to candidate policy families and evaluation requirements.}\label{fig:decision}
\end{figure}

The simulation is most useful when it changes the structure of a policy decision rather than when it produces a single percentage. Figure~\ref{fig:decision} translates the evidence and ABM mechanisms into a decision flow. If harm is concentrated at micro-places, place-based analysis and problem-oriented policing are the natural first diagnostic. If high-harm groups or networks drive severe violence, focused deterrence, investigation, and exit services become more relevant. If the mechanism is local compliance, illicit livelihood, or information held by legitimate community organizations, community-control or Guardabosques-type tools become plausible only after anti-capture and anti-vigilantism safeguards are satisfied. If risk is developmental or related to repeat offending, CBT, school retention, employment, treatment, and reentry are more coherent. If opportunity is generated by the physical environment, lighting, alcohol regulation, remediation, transport-node design, and municipal ``super-controller'' tools are appropriate candidates \parencite{shader2024,mazerolle2025partnerships}.

This decision logic matters particularly for Guardabosques-type programs. A community incentive should not pay residents simply for reporting fewer crimes, because that creates a direct incentive to suppress reporting. The relevant verified outcome must be one that the community can legitimately influence and that can be independently measured. Examples include environmental restoration, school retention, verified participation in legal economic alternatives, reduced recruitment into high-risk networks, or compliance with locally legitimate rules, while victimization is measured independently through surveys or external administrative sources. Confidential reporting, witness protection, anti-retaliation rules, independent auditing, and clear limits on community sanctioning are essential. Community institutions can generate information and compliance, but they should not become private coercive authorities.

A place-based policing intervention likewise requires safeguards. Hot-spots selection should be based on transparent concentration measures and should not be defined by demographic proxies. Outcomes should include victimization and severe harm rather than only stops, searches, or arrests. Adjacent spillover rings should be pre-specified so displacement can be measured. Procedural-justice and oversight indicators should be collected simultaneously. A local fall in recorded crime accompanied by lower reporting, higher complaints, or migration of harm to neighboring blocks should not be classified as an unqualified success.

Focused deterrence requires a credible identification of high-harm groups rather than broad labeling of neighborhoods or youth categories. The model therefore represents focused deterrence as a network/high-harm channel, not a demographic treatment. The policy should combine clear communication of consequences, services and exit opportunities, community messaging, and proportionate enforcement. The external literature suggests meaningful effects in targeted violence contexts, but the randomized subset is smaller than the pooled observational estimate \parencite{braga2026}. The ABM responds to this by using conservative prior means and uncertainty rather than mechanically applying the pooled percentage.

Organized crime poses another identification problem. Overall victimization surveys do not measure financial-network disruption, money laundering, precursor logistics, witness intimidation, or cross-border criminal organization capacity. The AML/network and investigation modules are therefore only weakly anchored by EH and are mostly structural/external-prior channels. Their evaluation should eventually incorporate case-clearance rates, financial intelligence, asset recovery, network fragmentation, judicial outcomes, witness safety, and homicide data. Until those data are linked, the model should not use a small overall victimization effect to conclude that investigative capacity is unimportant. Its primary outputs may lie elsewhere.

Hospital-linked violence intervention and violence interrupters illustrate a similar problem. These programs target selected high-risk populations and may affect reinjury, retaliation, or shootings rather than general property crime. The literature is heterogeneous and recent meta-analysis continues to emphasize uncertainty \parencite{webster2023hvip,east2026hvip}. A national-average ABM should therefore keep their reach low unless the synthetic population explicitly models emergency-department cohorts or street-group networks. The architecture supports such extensions but the included development results do not claim a separate causal estimate for HVIP.

Model failure can arise from policy saturation. A treatment with 100\% nominal intensity cannot automatically serve every eligible agent. Police units, prosecutors, community organizations, and social programs have finite capacity. The code includes capacity and budget ledgers so allocation cannot exceed configured limits. The current development scenarios use simplified relative capacities, while a production implementation should estimate police-hours, prosecutor caseload, CBT slots, program budgets, street-lighting unit costs, witness-protection capacity, and municipal staffing. Capacity constraints are not merely accounting details: they can reverse the ranking of portfolios when a high-intensity policy becomes impossible to implement with fidelity.

Capture and corruption are second-order mechanisms with first-order policy consequences. Community organizations can be captured by local political or criminal interests. Police and justice institutions can face corruption attempts or distorted information. Oversight and procedural quality reduce those risks in the conceptual architecture. The development model expresses capture as a scalar degradation of community effectiveness because the available surveys do not identify a richer corruption process. A production model should introduce network-based capture, strategic corruption attempts, internal-affairs detection, protected reporting, procurement risks, and judicial independence. Those mechanisms would be especially important in simulations of organized crime and border logistics.

Under-reporting is another failure mode. A low observed reporting rate means administrative crime counts are a noisy observation process rather than the latent outcome. The model therefore calibrates a reporting shift and preserves true simulated events separately from reported events. A reform that improves trust can increase recorded crime even when true victimization falls. This is not an inconsistency; it is a central reason to combine victimization surveys with police data. If a future administrative dataset is added, the likelihood should jointly model latent crime and the observation/reporting process instead of treating recorded incidents as error-free truth.

The evaluation strategy should be prospective. The model is not a substitute for field experimentation or quasi-experimental evaluation. It is a tool for selecting which combinations deserve piloting and which outcomes must be measured. For municipal rollouts, randomized or stepped-wedge assignment is appropriate where feasible. When randomization is impossible, matched difference-in-differences, synthetic controls, interrupted time series, regression discontinuity, or network designs can be used depending on the intervention. Pre-registration should specify primary outcomes, spillover zones, displacement measures, subgroup analyses, and multiple-testing corrections. Null effects should be retained rather than filtered from the evidence base.

A staged implementation can use the model recursively. First, calibrate baseline risk and institutional states from survey and administrative data. Second, screen broad packages in the ABM under deliberately wide uncertainty. Third, identify portfolios that remain acceptable under adverse adaptation, displacement, and fidelity draws. Fourth, pilot a small number of distinguishable packages in real territories. Fifth, update the emulator and priors with observed implementation data. Sixth, scale only if real-world outcomes are directionally consistent and no major rights or displacement failure appears. This creates a policy-learning loop rather than a one-time simulation exercise.

\section{Interpretation, limitations, and conclusions}

The strongest conclusion from the simulation is methodological rather than numerical. Crime reduction is a portfolio-design problem under uncertainty. Different mechanisms call for different institutions, time horizons, and outcome measures. Place-based policing can be highly influential for short-run victimization because harm is concentrated spatially and because the evidence base for targeted place interventions is relatively strong. Focused deterrence and investigation matter more when severe violence is concentrated in networks. Procedural justice matters primarily for legitimacy and reporting. Community social control and Guardabosques-type programs are plausible where legitimate organizations possess local information and where economic substitution changes incentives, but they are not credible substitutes for homicide investigation or financial intelligence. Social prevention is slower but affects recurrence and developmental risk. Environmental interventions alter opportunity. A coherent architecture must allow these mechanisms to complement rather than rhetorically replace one another.

The included development run is intentionally conservative about what can be learned from two surveys. ECE and EH provide a strong socioeconomic backbone and, unusually, EH supplies victimization, reporting, perceived safety, and police-trust indicators. They do not identify homicide dynamics, gang networks, extortion prevalence, judicial case processing, municipal policy capacity, weapons, money laundering, or organized-crime logistics. Those mechanisms are represented using external evidence and structural assumptions. The simulation therefore has more empirical leverage over baseline common victimization and reporting than over rare severe violence or organized crime.

Rare events are especially difficult. The development population contains only 5,000 synthetic persons and the scenario engine uses a severe-violence escalation process. Homicide counts at this scale are sparse, so point estimates would be unstable. The report emphasizes severe-violence rates and uncertainty rather than presenting precise homicide effects. Production analyses should use much larger synthetic populations, longer horizons, administrative homicide data, and rare-event calibration. The production configuration supports 50,000 agents in the included script, but that full national-scale run was not executed in the present environment and no claim is made that the development result is a converged production forecast.

The emulator design is another limitation. Gaussian processes provide smooth uncertainty-aware approximation but can struggle in high-dimensional or discontinuous response surfaces. The Bayesian-bootstrap tree ensemble provides a nonlinear complement but is not exact BART. Stacking reduces dependence on one surrogate, yet the sensitivity analysis remains conditional on the emulator training design. Active learning should be expanded around regions of emulator disagreement and high expected policy regret. Emulator diagnostics should also be computed separately for each outcome because an emulator adequate for common victimization may be inadequate for severe violence.

The sensitivity analysis demonstrates model structure rather than empirical causal decomposition. Hot-spots intensity dominates the short-run emulated victimization response because the ABM translates place-focused evidence into a comparatively direct opportunity channel. That result should trigger a validation question, not a political prescription: does real Bolivian micro-place data show comparable concentration, and can policing be targeted without excessive displacement or rights costs? The appropriate next step is to obtain geocoded incident and victimization information and test the concentration mechanism directly.

Policy-prior transportability is also uncertain. Meta-analyses combine studies from institutional environments that differ from Bolivia. Even Latin American studies differ in urban form, police organization, justice capacity, illicit markets, and reporting. The package partially addresses this by shrinking prior effects, adding uncertainty, and separating directness from credibility in the evidence review. A more formal transportability layer could model context moderators such as state capacity, baseline violence, police legitimacy, inequality, urban density, and implementation fidelity. Hierarchical evidence synthesis could then update intervention priors as Bolivian pilot data accumulate.

The community-governance results require particular care. The small simulated national-average effect of community control or Guardabosques does not contradict the evidence that such approaches can reduce conflict or support compliance in specific settings. It reflects limited assumed reach in the national-style model and the refusal to extrapolate a targeted governance mechanism to every crime category. In a coca-producing municipality, a Guardabosques-like incentive tied to legal livelihoods, environmental outcomes, and verified collective compliance could have much larger local effects on illicit-market recruitment or conflict. Conversely, a community program in a territory with high capture risk could perform worse than the same design in a cohesive, legitimate organization. The architecture is intended to test precisely that heterogeneity.

The distributional stress test is reassuring only in a narrow sense. The integrated package generates broadly similar simulated reductions across poor/non-poor, urban/rural, and youth/adult groups. That does not establish fairness. Equal reductions can coexist with unequal surveillance, use of force, arrest, incarceration, or exposure to coercive contacts. A production fairness module should therefore include enforcement exposure, false-positive targeting, complaint rates, detention, procedural outcomes, and program access by demographic and socioeconomic groups. Fairness cannot be inferred solely from victimization reductions.

The cost index is similarly preliminary. It measures relative resource intensity but not fiscal cost in Bolivianos. A realistic budget model should price police hours, vehicles, intelligence analysts, prosecutors, courts, treatment slots, streetlights, maintenance, municipal staff, grants, witness protection, employment subsidies, and data infrastructure. Cost-effectiveness should then incorporate discounted multi-year benefits, victimization costs, health costs, lost income, and distributional welfare. The current Pareto frontier should be understood as a structural demonstration of trade-offs, not a fiscal recommendation.

Despite these limitations, the model improves on single-policy reasoning in several ways. First, it forces all claims into explicit channels and prevents a policy from receiving benefits on outcomes it was not designed to affect. Second, it uses paired common random numbers so scenario differences are not dominated by Monte Carlo noise. Third, it propagates posterior, prior, and structural uncertainty rather than presenting one deterministic run. Fourth, it includes adaptation, displacement, capture, legitimacy, and reporting as endogenous or uncertain mechanisms. Fifth, it keeps empirical targets, external evidence, assumptions, calibration, and simulations separately labeled. Sixth, it provides complete code, tests, derived targets, configuration files, UML diagrams, and reproducibility commands so the analysis can be challenged and extended.

The principal numerical result is therefore best read as a proof of comparative behavior under the specified model. The full integrated package reduces simulated victimization by a paired mean of approximately 17.9\% relative to baseline in the development ensemble, while the place-based portfolio produces a similar 17.5\% paired reduction at lower resource cost but with a smaller severe-violence benefit. Focused policing produces about a 13.4\% paired victimization reduction. Enforcement-heavy policy produces substantial severe-violence reduction but the highest displacement among active portfolios. Rights-first and justice-legitimacy packages have small direct crime effects but large trust effects. Community-control and Guardabosques-type modules have small national-average crime effects while preserving their conceptual role in local compliance, legitimacy, and livelihood substitution. These patterns are coherent with the model's explicit mechanisms and evidence priors, but they remain simulated outcomes rather than causal policy forecasts.

The policy implication is not a ranking of political choices. It is a decision architecture. A government or municipality should begin by identifying whether harm is concentrated in places, networks, households, illicit markets, developmental trajectories, or opportunity structures. It should then select interventions whose mechanisms match that concentration, include safeguards appropriate to the institution, and evaluate real outcomes with designs capable of detecting displacement and unintended harm. The ABM can then be updated as new evidence arrives. In that sense, the model is not an answer machine; it is a disciplined way to state assumptions, compare combinations, expose trade-offs, and decide what must be learned empirically before scale-up.

The next empirical frontier is clear. Bolivia would benefit from a harmonized, privacy-preserving research infrastructure linking geocoded incidents, homicide and injury data, prosecutor and court trajectories, calls for service, victimization surveys, program participation, municipal environments, school outcomes, employment, business licensing, alcohol outlets, lighting, and financial-investigation outcomes. With those data, the model could move from a survey-anchored structural microsimulation toward a richer spatial-network calibration. Until then, the present package provides a transparent baseline: a path-dependent, multi-agent, multi-policy system in which uncertainty and failure modes are visible rather than hidden.

\printbibliography[title={References}]
\end{document}